\documentclass[]{raa}
\usepackage{amssymb}            
\usepackage{graphicx,times}
\usepackage{natbib}

\providecommand{\bm}[1]{\mbox{\boldmath$#1$\unboldmath}}

\def\Hbeta{\rm H{\beta}}
\def\Mgii{\rm Mg~ II}
\def\Civ{\rm C~ IV}

\begin{document}

   \title{SDSS spectroscopy for blazars in the $Fermi$ LAT bright AGN sample
}

 \volnopage{ {\bf 2009} Vol.\ {\bf 9} No. {\bf XX}, 000--000}
   \setcounter{page}{1}

   \author{Zhaoyu Chen
      \inst{1,2}
   \and Minfeng Gu
      \inst{1}
    \and Zhonghui Fan
      \inst{3}
    \and Xinwu Cao
      \inst{1}
   }

   \institute{Key Laboratory for Research in Galaxies and Cosmology,
Shanghai Astronomical Observatory, Chinese Academy of Sciences, 80
Nandan Road, Shanghai 200030, China; {\it zychen@shao.ac.cn}\\
        \and
             Graduate School of the Chinese Academy of Sciences, Beijing 100039, China
        \and
             Department of Physics, Yunnan University, Kunming 650091, Yunnan, China\\
\vs \no
   {\small Received [year] [month] [day]; accepted [year] [month] [day] }
}

\abstract{We have collected all available spectra and photometric
data from SDSS catalogue for bright AGNs complied from the first
three months of the $Fermi$ large area telescope all-sky survey.
Based on the 106 high-confidence and 11 low-confidence association
bright AGN list, the photometry data are collected from SDSS DR7 for
28 sources (12 BL Lacs and 16 FSRQs), two of which are
low-confidence association bright AGNs. Among these 28 SDSS
photometric sources, SDSS spectra are available for 20 sources (6 BL
Lacs and 14 FSRQs). The black hole mass $M_{\rm BH}$ and the broad
line region (BLR) luminosity were obtained for 14 FSRQs by measuring
the line-width and strength of broad emission lines from SDSS
spectra. The broad emission lines measurements of five FSRQs are
presented for the first time in this work. The optical continuum
emission of these 14 FSRQs is found to be likely dominated by the
nonthermal jet emission through comparing the relationship between
the broad Mg II line and continuum luminosity to that of radio quiet
AGNs. The black hole mass of 14 FSRQs ranges from $\rm
10^{8.2}M_{\odot}$ to $\rm 10^{9.9}M_{\odot}$, with most of sources
larger than $\rm 10^{9}M_{\odot}$. The Eddington ratio $L_{\rm
bol}/L_{\rm Edd}$ ranges from $10^{-1.5}$ to $\sim 1$. This implies
that the optically thin, geometrically thick accretion disk may
exist in these FSRQs. \keywords{galaxies: active
--- galaxies: nuclei
--- gamma rays: observations --- quasars: emission lines --- BL
Lacertae objects: general
 } }

   \authorrunning{Zhaoyu Chen, M. F. Gu, Z. Fan \& X. Cao}            
   \titlerunning{SDSS spectroscopy for $Fermi$-detected bright AGNs}  
   \maketitle


%
%
\section{Introduction}           
\label{sect:intro}

Blazars, including BL Lac objects and flat-spectrum radio quasars
(FSRQs), are the most extreme class of active galactic nuclei
(AGNs), characterized by strong and rapid variability, high
polarization, and apparent superluminal motion. These extreme
properties are generally interpreted as a consequence of non-thermal
emission from a relativistic jet oriented close to the line of
sight. As such, they represent a fortuitous natural laboratory with
which to study the physical properties of jets, and, ultimately, the
mechanisms of energy extraction from the central supermassive black
holes.

The most prominent characteristic of the overall spectral energy
distribution (SED) of blazars is the double-peak structure with two
broad spectral components. The first, lower frequency component is
generally interpreted as being due to synchrotron emission, and the
second, higher frequency one as being due to inverse Compton
emission. BL Lac objects (BL Lacs) usually have no or only very weak
emission lines, but have a strong highly variable and polarized
non-thermal continuum emission ranging from radio to ${\rm
\gamma}$-ray band, and their jets have synchrotron peak frequencies
ranging from IR/optical to UV/soft-X-ray energies. Compared to BL
Lacs, FSRQs have strong narrow and broad emission lines, however
generally have low synchrotron peak frequency. According to the
synchrotron peak frequency, BL Lac objects can be divided into three
subclasses, i.e. low frequency peaked BL Lac objects (LBL),
intermediate objects (IBL) and high frequency peaked BL Lac objects
(HBL) (Padovani \& Giommi 1995).

Although a point of agreement about the second component is that the
$\gamma$-rays of blazars are produced in relativistic jets by
inverse Compton scattering, the origin of the seed photons
(optical/IR), the location and size of the emitting region, and the
degree of relativistic beaming of the high-energy radiation, are all
still unknown. Seed photons may come from the jet itself
(Synchrotron Self-Compton model: SSC), from an accretion disk around
a supermassive black hole at the base of the jet, or else from
photons of the broad emission line region (e.g. Wehrle et al. 1998;
Collmar et al. 2000). 
If the $\gamma$-rays origins from up-scattering of broad emission
line photons, a correlation between the $\gamma$-rays and the
emission lines would be expected. However, a direct comparison shown
that there is no evidence of correlation between the $\gamma$-rays
and broad emission lines luminosity (Fan 2000), likely due to the
lack of line data. Through estimating the energy density in the
relativistic jet blobs contributed by the broad line region (BLR),
Fan, Cao \& Gu (2006) claimed that the $\gamma$-ray emission can be
from external Compton scattering, especially with seed photons from
broad line region. In the model fitting to SED including
$\gamma$-ray emission, the external Compton process have been taken
into account in combination with SSC (Ghisellini et al. 1998;
Ghisellini \& Tavecchio 2008). The measurements of broad emission
lines can be used to constrain the energy density of BLR in jets,
although it also depends on the location of jet region. On the other
hand, the measurements of prominent broad emission lines, especially
$\rm H\beta$, Mg II and C IV, enable us to measure the black hole
mass and the bolometric luminosity using various empirical
relations, from which the accretion mode can be investigated.

The $Gamma~ ray~ Large~ Area~ Space~ Telescope~ (GLAST)$ was
launched on 11 June 2008, and renamed the $Fermi~ Gamma~ Ray~ Space~
Telescope$ shortly after entering its scientific operating mission,
which began on 11 August, 2008. Abdo et al. (2009a) have presented a
list of 116 bright, $\gtrsim 10\sigma$ sources at
$|b|\geq10^{\circ}$ taken from the list of bright sources (Abdo et
al. 2009b) observed with the $Fermi$ Large Area Telescope (LAT) in
its initial three-month observing period extending from August 4 to
October 30 of 2008. Of these sources, 106 are associated with
blazars with high confidence and compose the LAT Bright AGN Sample
(LBAS). It contains two radio galaxies, namely Centaurus A and NGC
1275, and 104 blazars consisting of 57 FSRQs, 42 BL Lac objects, and
5 blazars with uncertain classification. The number of
low-confidence AGN associations is 11 (one source having two
possible associations - one high and one low confidence). In this
work, we search the SDSS photometry and spectroscopy data for all
these 116 AGNs, and the results focussing on the spectral line
analysis are shown. The cosmological parameters ${\rm H_0=70 \,
km~s^{-1}~Mpc^{-1}}$, $\Omega_{\rm m}$=0.3, $\Omega_{\Lambda}$ = 0.7
are used throughout the paper, and the spectral index $\alpha$ is
defined as $f_{\nu}$ $\propto$ $\nu^{-\alpha}$ with $f_{\nu}$ being
the flux density at frequency $\nu$.



\section{SDSS data analysis}
\label{sect:analysis}


The Sloan Digital Sky Survey (SDSS)\footnote{http://www.sdss.org}
consists of a series of three interlocking imaging and spectroscopic
surveys (Legacy, SEGUE, and Supernova), carried out over an
eight-year period with a dedicated 2.5m telescope located at Apache
Point Observatory in Southern New Mexico. The seventh data release
(DR7) from the SDSS represents the completion of this project. The
DR7 imaging data cover about 8423 square degrees of ``legacy" sky,
with information on roughly 230 million distinct photometric
objects, and about 3240 square degrees of SEGUE sky, with about 127
million distinct objects (including many stars at low latitude). The
DR7 spectroscopic data include data from 1802 main survey plates of
640 spectra each, and cover 8200 square degrees. In addition, DR7
contains 762 ``extra" and ``special" plates.

Based on the 106 high-confidence association and 11 low-confidence
association bright AGN list, the photometry data are collected from
SDSS DR7 for 28 sources (12 BL Lacs and 16 FSRQs), which is listed
in Table \ref{tbl1}. Two out of 28 objects are low-confidence
association bright AGNs, i.e. BL Lac object 0FGL J0909.7+0145
identified with SDSS J090939.84+020005.2 ($z=1.575$), and FSRQ 0FGL
J1641.4+3939 associated with SDSS J164147.54+393503.3 ($z=0.539$).
Among these 28 SDSS photometric sources, SDSS spectra are available
for 20 sources (6 BL Lacs and 14 FSRQs). In below, we present the
data analysis based on the available SDSS data.

\subsection{Photometry}

SDSS DR7 provides an imaging catalogue of 357 million unique
objects, from which the magnitude are available at $ugriz$ wavebands
with the average wavelengths $\rm 3551\AA, 4686\AA, 6165\AA,
7481\AA$ and $\rm 8931\AA$, respectively.
As blazars often appear as point-like sources (although the faint
host galaxy can be seen for BL Lac objects in some case), the PSF
magnitude from SDSS are collected for all 28 sources, which is
determined by fitting a PSF model to the object (see Table 2). The
SDSS $ugriz$ magnitudes were firstly transferred to AB $ugriz$
magnitudes using the zeropoint offset between the SDSS system and
the AB system (Oke \& Gunn 1983), i.e. $u_{\rm AB} = u_{\rm SDSS} -
0.04$ mag, $g_{\rm AB} \sim g_{\rm SDSS}$, $r_{\rm AB} \sim r_{\rm
SDSS}$, $i_{\rm AB} \sim i_{\rm SDSS}$ and $z_{\rm AB} = z_{\rm
SDSS} + 0.02$ mag. The Galactic extinction are then corrected using
the extinction law of Schlegel, Finkbeiner \& Davis (1998) and the
$E(B-V)$ provided in the NASA/IPAC Extragalactic Database
(NED)\footnote{http://nedwww.ipac.caltech.edu/}. Finally, flux
density can be calculated from magnitudes at each wavebands of
$ugriz$ as:
\begin{equation}
S_{\nu}=3631\times10^{\rm -0.4~m} \rm Jy
\end{equation}
in which $S_{\nu}$ and m are the flux density and Galactic
extinction corrected AB magnitude at $ugriz$ wavebands,
respectively. The simple power-law $S_{\nu}=a\nu^{-\alpha_{\nu}}$ is
used to fit the flux density at $ugriz$ wavebands. The spectral
index $\alpha_{\nu}$ is presented in Table \ref{tbl2}, in which the
luminosity at $\rm 5100\AA$ calculated from power-law fit is also
given.

\subsection{Spectral analysis}

The SDSS spectra cover the wavelength range from 3800 to $\rm
9200\AA$ with a resolution of about 1800 - 2200. In a first step,
the SDSS spectra for all 20 sources were corrected for the Galactic
extinction using the reddening map of Schlegel et al. (1998) and
then shifted to their rest wavelength, adopting the redshift (from
NED) listed in Table 1.

\subsubsection{BL Lac objects}

Historically, an object is classified as a BL Lac object based on a
flat radio spectrum ($\alpha_{\nu}\leq0.5$) and an optical spectrum
with emission lines weaker than $\rm 5~\AA$ rest-frame equivalent
width, e.g. for the 1-Jy radio survey (Stickel et al. 1991; Rector
\& Stocke 2001). The latter criterion was chosen to separate BL Lacs
from FSRQ. However, this classification arose controversy. 
Later on, the absorption feature Ca H\&K break was proposed in BL
Lacs classification in combination with the criteria of no or very
weak emission lines in the optical spectrum (March\~{a} et al. 1996;
see also Landt, Padovani \& Giommi 2002). In this work, we follow
the BL lacs identification in Abdo et al. (2009a) (see their Tables
1 and 2).

The SDSS spectra of 6 BL Lacs are featureless, and the Ca H\&K break
is not evident. Therefore, we use a single power-law to fit the
spectral continuum, and the contribution of host galaxy is ignored.
The spectral index is shown in Table 3, as well as the luminosity at
$\rm 5100\AA$ calculated from the fitted power-law.


\subsubsection{FSRQs}
The spectra of quasars are characterized by various of broad and
narrow emission lines (Vanden Berk et al. 2001). The redshift of our
14 FSRQs ranges from 0.539 to 2.189 (see Table 1), with several
broad emission lines apparently seen from SDSS spectra, e.g. $\rm
H\beta$, Mg II and C IV lines (see Fig. 1). For our FSRQs sample, we
ignore the host galaxy contribution to the spectrum, since only very
little, if any, starlight is observed. In order to reliably measure
line parameters, we choose those wavelength ranges as
pseudo-continua, which are not affected by prominent emission lines,
and then decompose the spectra into the following three components
(see also Chen, Gu \& Cao 2009):

1. A power-law continuum to describe the emission from the active
nucleus. The 13 line-free spectral regions were firstly selected
from SDSS spectra covering 1140$\rm \AA$ to 5630$\rm \AA$ for our
sample, namely, 1140$\rm \AA$ -- 1150$\rm \AA$, 1275$\rm \AA$ --
1280$\rm \AA$, 1320$\rm \AA$ -- 1330$\rm \AA$, 1455$\rm \AA$ --
1470$\rm \AA$, 1690$\rm \AA$ -- 1700$\rm \AA$, 2160$\rm \AA$ --
2180$\rm \AA$, 2225$\rm \AA$ -- 2250$\rm \AA$, 3010$\rm \AA$ --
3040$\rm \AA$, 3240$\rm \AA$ -- 3270$\rm \AA$, 3790$\rm \AA$ --
3810$\rm \AA$, 4210$\rm \AA$ -- 4230$\rm \AA$, 5080$\rm \AA$ --
5100$\rm \AA$, 5600$\rm \AA$ -- 5630$\rm \AA$ (Vanden Berk et al.
2001; Forster et al. 2001). Depending on the source redshift, the
spectrum of individual quasar only covers some of them, from which
the initial single power-law are obtained for each source. 

2. An Fe II template. The spectra of our sample covers UV and
optical regions, therefore, we adopt the UV Fe II template from
Vestergaard \& Wilkes (2001), and optical one from V\'{e}ron-Cetty
et al. (2004).
For the sources with both UV Fe II and optical Fe II lines prominent
in the spectra, we connect the UV and optical templates into one
template covering the whole spectra (see also Chen et al. 2009). In
the fitting, we assume that Fe II has the same profile as the
relevant broad lines, i.e. the Fe II line width usually was fixed to
the line width of broad $\Hbeta$ or $\Mgii$ or $\Civ$, which in most
cases gave a satisfied fit. In some special cases, a free-varying
line width was adopted in the fitting to get better fits.

3. A Balmer continuum generated in the same way as Dietrich et al.
(2002) (see also Chen et al. 2009). Grandi (1982) and Dietrich et
al. (2002) proposed that a partially optically thick cloud with a
uniform temperature could produce the Balmer continuum, which can be
expressed as:
\begin{equation}
F_{\lambda}^{\rm BaC}=F_{\rm
BE}B_{\lambda}(T_{e})(1-e^{-\tau_{\lambda}}); \qquad (\lambda<
\lambda_{\rm BE}) \label{Flamd.eq}
\end{equation}
where $F_{\rm BE}$ is a normalized coefficient for the flux at the
Balmer edge $(\lambda_{\rm BE}=3646\rm \AA)$, $B_{\lambda}(T_{e})$
is the Planck function at an electron temperature $T_e$, and
$\tau_{\lambda}$ is the optical depth at
 $\lambda$ and is expressed as:
\begin{equation}
\tau_{\lambda}=\tau_{\rm BE}(\frac{\lambda}{\lambda_{\rm BE}})
\label{taulamd.eq}
\end{equation}
where $\tau_{\rm BE}$ is the optical depth at the Balmer edge. There
are two free parameters, $F_{\rm BE}$ and $\tau_{\rm BE}$. Following
Dietrich et al. (2002), we adopt the electron temperature to be $T_e
\, =\, 15,000$ K.

The modeling of above three components is performed by minimizing
the $\chi^2$ in the fitting process. The final multicomponent fit is
then subtracted from the observed spectrum. The fitted power-law, Fe
II lines, Balmer continuum and the residual spectra for each source
are shown in Fig. \ref{fig1}. The Fe II fitting windows are selected
as the regions with prominent Fe II line emission while no other
strong emission lines, according to Vestergaard \& Wilkes (2001) and
Kim et al. (2006). The fitting window around the Balmer edge ($3625
- 3645\rm \AA$) is used to measure the contribution of Balmer
continuum, which extends to $\Mgii$ line region. The Balmer
continuum is not considered when $3625 - 3645\rm \AA$ is out of the
spectrum. 

The broad emission lines were measured from the continuum subtracted
spectra. We mainly focused on several prominent emission lines, i.e.
$\Hbeta$, Mg II, C IV. Generally, two gaussian components were
adopted to fit each of these lines, indicating the broad and narrow
line components, respectively. The blended narrow lines, e.g. [O
III] $\lambda\lambda4959,5007\rm \AA$ and [He II] $\rm \lambda 4686
\AA$ blending with $\rm H\beta$ were included as one gaussian
component for each line at the fixed line wavelength. The $\chi^{2}$
minimization method was used in fits. The line width FWHM, line flux
of broad $\Hbeta$, Mg II and C IV lines were obtained from the final
fits for our sample, and are listed in Table 4. The spectral fitting
for each source are shown in Fig. \ref{fig1}.


\subsection{$M_{\rm bh}$ and $L_{\rm BLR}$} \label{mbhlbol}

There are various empirical relations between the radius of broad
line region (BLR) and the continuum luminosity, which can be used to
calculate the black hole mass in combination with the line width
FWHM of broad emission lines. However, there are defects when using
the continuum luminosity to estimate the BLR radius for blazars
since the continuum flux of blazars are usually Doppler boosted due
to the fact that the relativistic jet is oriented close to the line
of sight. Alternatively, the broad line emission can be a good
indicator of thermal emission from accretion process. Therefore for
our FSRQs sample, we estimate the black hole mass by using the
empirical relation based on the luminosity and FWHM of broad
emission lines. According to the source redshift, we use various
relations to estimate the black hole mass: Vestergaard \& Peterson
(2006) for broad $\rm H\beta$ (see also Wu et al. 2004) and Kong et
al. (2006) for broad Mg II and C IV lines.

For the sources with available FWHM and luminosity of broad
$\Hbeta$, the method to calculate $M_{\rm BH}$ is given by
Vestergaard \& Peterson (2006):
\begin{equation}
M_{\rm BH} (\Hbeta) = 4.68\times10^6 
   \left( \frac{\it L(\rm H\beta)}{10^{42}\, \rm erg~s^{-1}}\right)^{0.63}
  \left(\frac{\rm FWHM(H\beta)}{1000~km~s^{-1}} \right)^2 {\it M}_{\sun}
\label{MBH_LHb.eq}
\end{equation}

In addition, Kong et al. (2006) presented the empirical formula to
obtain the black hole mass using broad $\Mgii$ and $\Civ$ for high
redshift sources as follows,
\begin{equation}
  M_{\rm BH}({\rm Mg~II}) = 2.9 \times 10^{6} \left
  (\frac{L({\rm Mg~II})}{\rm 10 ^{42}\, erg\, s^{-1}}\right )^{0.57\pm0.12}
  \left(\frac{\rm FWHM(Mg~II)}{\rm 1000\, km\, s^{-1}}\right )^{2}
  M_{\sun},
\label{MBH_Mgii.eq}
\end{equation}

\begin{equation}
  M_{\rm BH}({\rm C~IV}) = 4.6 \times 10^{5} \left
  (\frac{L({\rm C~IV})}{\rm 10 ^{42}\, erg\, s^{-1}}\right )^{0.60\pm0.16}
  \left(\frac{\rm FWHM(C~IV)}{\rm 1000\, km\, s^{-1}}\right )^{2} M_{\sun}
\label{MBH_Civ.eq}
\end{equation}

In the redshift range of our 14 FSRQs, $M_{\rm BH}$ can be estimated
using two of above relations for 7 sources, in which two $M_{\rm
BH}$ values are consistent with each other within a factor of three.
Therefore, we adopted the average value of two estimates.

In this work, we estimate the BLR luminosity $L_{\rm BLR}$ following
Celotti et al. (1997) by scaling the strong broad emission lines
$\Hbeta$, Mg II and C IV to the quasar template spectrum of Francis
et al. (1991), in which Ly$\alpha$ is used as a reference of 100. By
adding the contribution of $\rm H\alpha$ with a value of 77, the
total relative BLR flux is 555.77, of which $\Hbeta$ 22, $\Mgii$ 34,
and $\Civ$ 63 (Celotti et al. 1997; Francis et al. 1991). From the
BLR luminosity, we estimate the bolometric luminosity as $L_{\rm
bol}=10L_{\rm BLR}$ (Netzer 1990). The black hole mass, BLR
luminosity and the Eddington ratio $L_{\rm bol}/L_{\rm Edd}$ are
shown in Table 5.

\section{Results and Discussion}
\label{sect:discussion}

In Abdo et al. (2009a), the combination of the figure of merit
approach (FoM) and positional association methods yields a number of
106 high-confidence (probability $P\geq0.90$) associations
(constituting the LBAS) and 11 low confidence ($0.40 < P < 0.90$)
associations. These 11 low-confidence association Bright AGNs
consist of 3 FSRQs, 2 BL Lacs, and 6 sources with uncertain
classification, of which all have a flat radio spectrum but there
are no other data reported in literature allowing to classify them
either as FSRQs or BL Lacs. For $\gamma$-ray source 0FGL
J1034.0+6051, two radio associations were found by the FoM method,
one with very high probability ($P=0.99$) and one with lower, but
still significant, probability ($P=0.79$). Although the
high-probability source likely dominates the $\gamma$-ray emission,
it is entirely plausible that the low-probability source contributes
non-negligibly to the total $\gamma$-ray flux (Abdo et al. 2009a).
In this work, the SDSS data is only available for the
high-confidence association (see Table 1), from which the emission
line measurements are given in Table 4 (also see Table 5). As stated
in Abdo et al. (2009a), the early results from the first three
months of the science mission of the $Fermi$ Gamma ray Space
Telescope demonstrate its exceptional capabilities to provide
important new knowledge about $\gamma$-ray emission from AGNs and
blazars. As the $Fermi$-LAT data accumulate, many more AGNs at lower
flux levels will likely be detected helping to improve our
understanding of supermassive black holes (Abdo et al. 2009a). As
predicted from AGN feedback by Wang (2008), there is extended faint
gamma-ray fuzz around radio-quiet quasars, or red quasars. $Fermi$
Gamma ray Space Telescope is anticipated to explore the fuzz more
feasibly. There is no doubt that the detection of the fuzz will
provide the most powerful diagnostic to the evolutionary chains of
galaxies and quasars from details of radiative feedback (see Wang
2008).

Radio-loud quasars, particularly FSRQs, differ from radio quiet AGNs
in the contribution of non-thermal jet emission in optical continuum
emission, in addition to the thermal emission from accretion disk.
Especially, the jet emission can be dominant due to the beaming
effect when jet is moving towards us with small viewing angle.
Indeed, Liu, Jiang \& Gu (2006) found that the $\rm 5100\AA$
continuum luminosity for FSRQs generally exceed that expected from
the relationship between the $\rm 5100\AA$ continuum and broad $\rm
H\beta$ luminosity for radio quiet AGNs (see also Gu, Chen \& Cao
2009). To inspect our FSRQs in a similar way, we investigate the
relationship between the $\rm 3000\AA$ continuum and broad Mg II
line luminosity, since Mg II line is available and measured for most
sources. The calibration of this relation is firstly performed for a
sample of radio quiet AGNs with reverberation mapping black hole
mass (Kaspi et al. 2000), by collecting the relevant data in Kong et
al. (2006). The ordinary least-square bisector linear fit gives:
\begin{equation}\label{ml30}
\rm log~ \it \lambda L_{\rm \lambda3000\AA}\rm = (0.996\pm0.048)~\rm
log~ \it L_{\rm Mg~ II}\rm +(1.895\pm2.024)
\end{equation}
which is plotted in Fig. 2, as well as the data points of radio
quiet AGNs.

The relationship between the broad Mg II luminosity and luminosity
at $\rm 3000\AA$ for our FSRQs is shown in Fig. 2. When Mg II line
is not available, we calibrate C IV line to Mg II line adopting the
relative flux (Mg II=34, C IV=63) of the relevant lines of the
composite spectrum of Francis et al. (1991). The luminosity at $\rm
3000\AA$ is calculated from the power-law fit on spectral continuum.
From Fig. 2, it can be seen that the $\rm 3000\AA$ luminosity of
most sources lies above the $\lambda L_{\lambda\rm 3000\AA} - L_{\rm
Mg ~II}$ relation of radio quiet AGNs. However, the deviation of
$\lambda L_{\lambda \rm 3000\AA}$ is within one order of magnitude.
The average value of $\lambda L_{\lambda \rm 3000\AA}/L_{\rm Mg~
II}$ is $\langle \rm log~\it (\lambda L_{\lambda \rm 3000\AA}/L_{\rm
Mg~ II})\rangle \rm =1.73\pm0.24$ for radio quiet AGNs, while it is
$\langle \rm log~\it (\lambda L_{\lambda \rm 3000\AA}/L_{\rm Mg~
II})\rangle \rm =2.21\pm0.27$ for our FSRQs. Therefore, the $\lambda
L_{\lambda \rm 3000\AA}$ of our FSRQs is, on average, higher than
that of radio quiet AGNs by a factor of three at given $L_{\rm Mg~
II}$, which is most likely due to the non-thermal jet emission. As
expected for FSRQs, this implies that the nonthermal jet emission in
FSRQs generally can be dominant over thermal emission, with the
former likely being Doppler boosted. Adopting a Doppler factor of
$\delta=10$ typical for FSRQs (Ghisellini et al. 1998; Jiang, Cao \&
Hong 1998; Gu, Cao \& Jiang 2009) and assuming the corresponding
Doppler boosting $\delta^{2+\alpha}$ (corresponding to a continuous
jet, $\alpha$ is the spectral index) in jet emission, the intrinsic
jet emission in optical band, however, can be much smaller than the
thermal emission (mainly from accretion disk) for our FSRQs.
However, the dependence of Doppler factor on the frequency was
proposed (Zhang, Fan \& Cheng 2002), with a lower Doppler factor
$\sim5$ claimed at optical region. In contrast, Chen et al. (2009)
shown that the thermal emission from accretion disk and host galaxy
can be dominant in 67 out of 185 FSRQs ($\sim36\%$) selected from
SDSS DR 3 quasar catalogue. As Landt et al. (2008) claimed, the flat
radio spectrum does not prove the jet-dominance in continuum
emission. The high radio core-dominance, in addition to flat radio
spectrum, are proposed to find jet-dominant FSRQs. We note that the
$\lambda L_{\lambda\rm 3000\AA} - L_{\rm Mg ~II}$ relation of
radio-quiet AGNs is established from relatively low luminosity,
which does not cover the luminosity range of our FSRQs. The further
investigations are needed.

It can be seen from Table \ref{tbl5} that the black hole masses of
14 FSRQs ranges from $\rm 10^{8.2}M_{\odot}$ to $\rm
10^{9.9}M_{\odot}$, with most of sources larger than $\rm
10^{9}M_{\odot}$. The BLR luminosity varies from $\rm 10^{44.44} erg
~s^{-1}$ to $\rm 10^{47.04} erg ~s^{-1}$. The Eddington ratio
$L_{\rm bol}/L_{\rm Edd}$ ranges from $10^{-1.5}$ to $\sim 1$. This
implies that the optically thin, geometric thick accretion disk
(Shakura \& Sunyaev 1973) may exist in these FSRQs. This is
consistent with the radio-loud quasars in Wang, Ho \& Staubert
(2003), of which the accretion rates, $\dot{M}\approx 0.01 - 1$
times the Eddington rate, suggesting that most of sources possess
standard optically thick, geometrically thin accretion disks. In
contrast, the accretion mode in BL Lacs is not homogenous, with
standard thin disks probably in luminous BL Lacs, and the
advection-dominated accretion flow in less luminous ones (e.g. Cao
2002). Moreover, the evolutionary sequence of blazars, from FSRQs to
low-energy-peaked BL Lac object to high-energy-peaked BL Lac object,
is suggested along the decreasing accretion rate (e.g. Cao 2002;
Cavaliere \& D'Elia 2002; Wang et al. 2003). While we estimate the
BLR luminosity from multiple emission lines following Celotti et al.
(1997), it is argued that the BLR luminosity can be better derived
from single line than multiple lines due to the very large scatter
in the ratio of some key lines (e.g. C IV and Mg II) with respect to
Ly $\alpha$ (Wang, Luo \& Ho 2004). Following the suggestion of Wang
et al. (2004), for seven of our FSRQs with multiple lines
measurements (see Table 4), we re-estimated the BLR luminosity using
only $\rm H\beta$ line, or Mg II when $\rm H\beta$ is not available,
which is listed in Table 5. We find that the BLR luminosity from
single line is all consistent with that from multiple lines within a
factor of two, except for SDSS J131028.66+322043.7 with the former
smaller than the latter by a factor of 4.7 (see Table 5). This
implies that the relative flux of $\rm H\beta$, Mg II and C IV, at
least for these seven FSRQs, generally follow the Francis et al.
(1991) composite spectrum, although they are $\gamma$-ray and strong
radio sources. Apparently, using single line does not significantly
alter the Eddington ratio estimates.

Our BLR measurements can be used in the further investigations
relevant to $Fermi~ \gamma$-ray emission. For example, the model fit
to SEDs including external Compton process and SSC will be present
for LBAS sources in a future paper (Li et al., in preparation), in
which the BLR measurements in this paper can be used to constrain
the energy density of BLR emission in jets. Moreover, to investigate
the mechanism of $\gamma$-ray emission, the work of Fan et al.
(2006), will be re-investigated for LBAS sources by using our BLR
measurements and collecting the BLR measurements from literatures
(Fan et al., in preparation). As a matter of fact, the BLR data for
five FSRQs (SDSS J082447.24+555242.6, SDSS J094857.33+002225.5, SDSS
J101603.14+051302.3, SDSS J155332.69+125651.7 and SDSS
J222940.09-083254.5) are not available from literatures, therefore,
are measured and provided in the first time in this paper (see Table
4). Despite extensive literature search, the BLR measurements are
still not available for some LBAS sources, therefore, the
spectroscopy observations are needed since the external Compton
process from BLR seed photons are thought to be primary for
$\gamma$-ray emission (Fan et al. 2006).


\normalem
\begin{acknowledgements}

We thank the anonymous referee for insightful comments and
constructive suggestions. This work is supported by National Science
Foundation of China (grants 10633010, 10703009, 10833002, 10773020
and 10821302), 973 Program (No. 2009CB824800), and the CAS
(KJCX2-YW-T03). ZHF is supported by Yunnan Provincial Science
Foundation of China (grant 2008CD061). This research made use of the
NASA/ IPAC Extragalactic Database (NED), which is operated by the
Jet Propulsion Laboratory, California Institute of Technology, under
contract with the National Aeronautics and Space Administration.

Funding for the SDSS and SDSS-II has been provided by the Alfred P.
Sloan Foundation, the Participating Institutions, the National
Science Foundation, the U.S. Department of Energy, the National
Aeronautics and Space Administration, the Japanese Monbukagakusho,
the Max Planck Society, and the Higher Education Funding Council for
England. The SDSS Web Site is http://www.sdss.org/.

The SDSS is managed by the Astrophysical Research Consortium for the
Participating Institutions. The Participating Institutions are the
American Museum of Natural History, Astrophysical Institute Potsdam,
University of Basel, University of Cambridge, Case Western Reserve
University, University of Chicago, Drexel University, Fermilab, the
Institute for Advanced Study, the Japan Participation Group, Johns
Hopkins University, the Joint Institute for Nuclear Astrophysics,
the Kavli Institute for Particle Astrophysics and Cosmology, the
Korean Scientist Group, the Chinese Academy of Sciences (LAMOST),
Los Alamos National Laboratory, the Max-Planck-Institute for
Astronomy (MPIA), the Max-Planck-Institute for Astrophysics (MPA),
New Mexico State University, Ohio State University, University of
Pittsburgh, University of Portsmouth, Princeton University, the
United States Naval Observatory, and the University of Washington.
\end{acknowledgements}



%

\begin{table}\bc
\begin{minipage}[]{100mm}
\caption[]{$Fermi$-detected AGNs in SDSS}\label{tbl1}\end{minipage}
\small
 \begin{tabular}{lccccc}
  \hline\noalign{\smallskip}
LAT Name & Other Name & SDSS Name &  Class  &
$z$   & SDSS data \\
(1)&(2)&(3)&(4)&(5)&(6)\\
  \hline\noalign{\smallskip}
0FGL J0050.5-0928 &   PKS 0048-097 & J005041.31-092905.1 & BLLac & 0.537 &  PS \\
0FGL J0738.2+1738 &   PKS 0735+178 & J073807.39+174218.9 & BLLac & 0.424 &  P \\
0FGL J0818.3+4222 &        OJ +425 & J081815.99+422245.4 & BLLac & 0.530 &  PS \\
0FGL J0824.9+5551 &   TXS 0820+560 & J082447.24+555242.6 &  FSRQ & 1.417 &  PS \\
0FGL J0909.7+0145$^{L}$ &   PKS 0907+022 & J090939.84+020005.2 &  BLLac & 1.575 &  PS \\
0FGL J0921.2+4437 &  RGB J0920+446 & J092058.46+444154.0 &  FSRQ & 2.190 &  PS \\
0FGL J0948.3+0019 & PMN J0948+0022 & J094857.33+002225.5 &  FSRQ & 0.585 &  PS \\
0FGL J0957.6+5522 &      4C +55.17 & J095738.19+552257.7 &  FSRQ & 0.896 &  PS \\
0FGL J1015.2+4927 &   1ES 1011+496 & J101504.14+492600.6 & BLLac & 0.212 &  PS \\
0FGL J1015.9+0515 & PMN J1016+0512 & J101603.14+051302.3 &  FSRQ & 1.715 &  PS \\
0FGL J1034.0+6051 &     S4 1030+61 & J103351.42+605107.3 &  FSRQ & 1.401 &  PS \\
0FGL J1057.8+0138 &   PKS 1055+018 & J105829.60+013358.8 &  FSRQ & 0.888 &  PS \\
0FGL J1058.9+5629 & RX J10586+5628 & J105837.73+562811.1 & BLLac & 0.143 &  PS \\
0FGL J1104.5+3811 &        Mrk 421 & J110427.31+381231.7 & BLLac & 0.030 &  P \\
0FGL J1159.2+2912 &      4C +29.45 & J115931.84+291443.8 &  FSRQ & 0.724 &  PS \\
0FGL J1218.0+3006 &     B2 1215+30 & J121752.08+300700.6 & BLLac & 0.130 &  P \\
0FGL J1221.7+2814 &          W Com & J122131.69+281358.4 & BLLac & 0.102 &  PS \\
0FGL J1229.1+0202 &         3C 273 & J122906.69+020308.5 &  FSRQ & 0.158 &  P \\
0FGL J1310.6+3220 &     B2 1308+32 & J131028.66+322043.7 &  FSRQ & 0.997 &  PS \\
0FGL J1427.1+2347 &   PKS 1424+240 & J142700.39+234800.0 & BLLac & ...   &  P \\
0FGL J1504.4+1030 &   PKS 1502+106 & J150424.98+102939.1 &  FSRQ & 1.839 &  PS \\
0FGL J1522.2+3143 &   TXS 1520+319 & J152209.99+314414.3 &  FSRQ & 1.487 &  P \\
0FGL J1553.4+1255 &   PKS 1551+130 & J155332.69+125651.7 &  FSRQ & 1.290 &  PS \\
0FGL J1555.8+1110 &     PG 1553+11 & J155543.04+111124.3 & BLLac & 0.360 &  P \\
0FGL J1635.2+3809 &      4C +38.41 & J163515.50+380804.4 &  FSRQ & 1.813 &  PS \\
0FGL J1641.4+3939$^{L}$ &    B3 1640+396 & J164147.54+393503.3 &  FSRQ & 0.539 &  PS \\
0FGL J1653.9+3946 &        Mrk 501 & J165352.21+394536.6 & BLLac & 0.033 &  P \\
0FGL J2229.8-0829 &       PHL 5225 & J222940.09-083254.5 &  FSRQ & 1.560 &  PS \\
  \noalign{\smallskip}\hline
\end{tabular}
\ec
\tablecomments{0.86\textwidth}{Col. (1): source LAT name. $^{L}$
represents low-confidence association AGNs. Col. (2): source other
name. Col. (3): source SDSS name. Col. (4): class. Col. (5):
redshift from NED. Col. (6): available SDSS data --- P for
photometry and S for spectroscopy.}
\end{table}

\begin{table}\bc
\begin{minipage}[]{100mm}
\caption[]{SDSS photometry}\label{tbl2}\end{minipage} \small
 \begin{tabular}{lcccccccc}
  \hline\noalign{\smallskip}
SDSS Name & $u$ & $g$ &  $r$  &
$i$   & $z$ & $E(B-V)$ & $\alpha_{\nu}$ & log $L_{\rm 5100 \AA}$ \\
&&&&&&&& $(\rm erg~s^{-1})$\\
(1)&(2)&(3)&(4)&(5)&(6)&(7)&(8)&(9)\\
  \hline\noalign{\smallskip}
J005041.31-092905.1  & 16.77 & 16.39 & 16.05 & 15.80 & 15.57 & 0.032 & 1.03 & 45.93 \\
J073807.39+174218.9  & 16.45 & 15.91 & 15.49 & 15.17 & 14.88 & 0.035 & 1.38 & 45.93 \\
J081815.99+422245.4  & 19.20 & 18.61 & 18.08 & 17.71 & 17.38 & 0.063 & 1.55 & 45.18 \\
J082447.24+555242.6  & 18.52 & 18.36 & 18.08 & 17.92 & 17.88 & 0.063 & 0.43 & 45.99 \\
J090939.84+020005.2$^{L}$  & 20.37 & 19.88 & 19.53 & 19.18 & 18.98 & 0.031 & 1.24 & 45.76 \\
J092058.46+444154.0  & 18.68 & 18.06 & 17.91 & 17.72 & 17.42 & 0.021 & 1.02 & 46.68 \\
J094857.33+002225.5  & 18.96 & 18.60 & 18.44 & 18.19 & 18.18 & 0.079 & 0.48 & 45.05 \\
J095738.19+552257.7  & 18.26 & 17.92 & 17.57 & 17.38 & 17.20 & 0.009 & 0.99 & 45.81 \\
J101504.14+492600.6  & 15.71 & 15.43 & 15.23 & 15.11 & 14.96 & 0.012 & 0.64 & 45.28 \\
J101603.14+051302.3  & 20.50 & 19.88 & 19.48 & 19.07 & 18.75 & 0.026 & 1.58 & 46.00 \\
J103351.42+605107.3  & 19.31 & 18.88 & 18.46 & 18.17 & 17.93 & 0.010 & 1.30 & 46.04 \\
J105829.60+013358.8  & 18.84 & 18.30 & 17.85 & 17.47 & 17.12 & 0.027 & 1.56 & 45.85 \\
J105837.73+562811.1  & 16.92 & 16.50 & 16.00 & 15.73 & 15.49 & 0.012 & 1.36 & 44.57 \\
J110427.31+381231.7  & 14.80 & 13.81 & 13.10 & 12.81 & 12.54 & 0.015 & 2.14 & 44.19 \\
J115931.84+291443.8  & 18.75 & 18.21 & 18.08 & 17.95 & 17.68 & 0.019 & 0.85 & 45.39 \\
J121752.08+300700.6  & 15.92 & 15.49 & 15.14 & 14.92 & 14.76 & 0.024 & 1.03 & 44.85 \\
J122131.69+281358.4  & 15.84 & 15.39 & 15.05 & 14.78 & 14.58 & 0.023 & 1.13 & 44.67 \\
J122906.69+020308.5  & 13.87 & 12.99 & 12.88 & 12.64 & 13.24 & 0.021 & 0.61 & 45.88 \\
J131028.66+322043.7  & 18.40 & 18.02 & 17.61 & 17.34 & 17.12 & 0.014 & 1.20 & 45.97 \\
J142700.39+234800.0  & 15.25 & 14.87 & 14.55 & 14.31 & 14.12 & 0.059 & 0.88 &  ... \\
J150424.98+102939.1  & 18.90 & 18.52 & 18.22 & 17.84 & 17.59 & 0.032 & 1.15 & 46.45 \\
J152209.99+314414.3  & 20.62 & 20.18 & 19.79 & 19.51 & 19.43 & 0.024 & 1.10 & 45.52 \\
J155332.69+125651.7  & 17.50 & 17.40 & 17.24 & 17.20 & 17.17 & 0.042 & 0.16 & 46.14 \\
J155543.04+111124.3  & 15.08 & 14.74 & 15.40 & 14.13 & 13.94 & 0.052 & 0.84 & 46.10 \\
J163515.50+380804.4  & 17.77 & 17.67 & 17.63 & 17.38 & 17.26 & 0.011 & 0.42 & 46.40 \\
J164147.54+393503.3$^{L}$  & 20.09 & 19.65 & 19.64 & 19.23 & 19.16 & 0.014 & 0.81 & 44.50 \\
J165352.21+394536.6  & 15.36 & 13.85 & 13.04 & 12.62 & 12.32 & 0.019 & 2.86 & 44.24 \\
J222940.09-083254.5  & 17.87 & 17.44 & 17.07 & 16.74 & 16.48 & 0.051 & 1.16 & 46.74 \\
  \noalign{\smallskip}\hline
\end{tabular}
\ec
\tablecomments{0.86\textwidth}{Col. (1): SDSS name. $^{L}$
represents low-confidence association AGNs. Cols. (2) - (6): SDSS
five-band PSF magnitude $ugriz$. Col. (7): $E(B-V)$ from NED. Col.
(8): the spectral index $\alpha_{\nu}$ from power-law fit on
$u,g,r,i,z$. Col. (9): luminosity at $\rm 5100\AA$.}
\end{table}

\begin{table}\bc
\begin{minipage}[]{100mm}
\caption[]{BL Lacs with SDSS spectra}\label{tbl3}\end{minipage}
\small
 \begin{tabular}{lccc}
  \hline\noalign{\smallskip}
SDSS Name &    $\alpha_{\lambda}$ & log $L_{\rm 5100 \AA}$ \\
 &  &  $(\rm erg~s^{-1})$\\
 (1)&(2)&(3)\\
  \hline\noalign{\smallskip}
J005041.31-092905.1       &    0.69 & 46.19  \\
J081815.99+422245.4       &    0.79 & 45.00  \\
J090939.84+020005.2$^{L}$ &    1.05 & 46.63  \\
J101504.14+492600.6       &    1.32 & 45.37  \\
J105837.73+562811.1       &    0.87 & 44.69  \\
J122131.69+281358.4       &    0.86 & 44.85  \\
  \noalign{\smallskip}\hline
\end{tabular}
\ec
\tablecomments{0.86\textwidth}{Col. (1): SDSS name. $^{L}$
represents low-confidence association AGNs. Col. (2): the spectral
index $\alpha_{\lambda}$ from power-law fit to SDSS spectra. Col.
(3): luminosity at $\rm 5100\AA$.}
\end{table}

\begin{table}\bc
\begin{minipage}[]{100mm}
\caption[]{Broad emission line measurements for
FSRQs}\label{tbl4}\end{minipage} \small
 \begin{tabular}{lcccccccccc}
  \hline\noalign{\smallskip}
SDSS Name   & C IV & & Mg II &   & $\rm H\beta$ & &
$\alpha_{\lambda}$ & log $L_{\rm 3000 \AA}$
\\
    &  FWHM  &log~$F$ & FWHM & log~$F$ & FWHM & log~$F$ \\
   & $(\rm km~s^{-1})$ &  &$(\rm km~s^{-1})$ & &$(\rm km~s^{-1})$ & & &   $(\rm erg~s^{-1})$\\
   (1) & (2) & (3) & (4) & (5) & (6) & (7) & (8) & (9)  \\
  \hline\noalign{\smallskip}
J082447.24+555242.6$^{*}$  &   ...  &     ... & 5119  &  -13.27 &  ...  &     ... & 1.39 &  46.88 \\
J092058.46+444154.0  &  8788  &  -12.68 &  ...  &     ... &  ...  &     ... & 1.18 &  47.69 \\
J094857.33+002225.5$^{*}$  &   ...  &     ... & 4743  &  -13.89 & 2209  &  -14.15 & 1.28 &  45.50 \\
J095738.19+552257.7  &   ...  &     ... & 6709  &  -13.70 &  ...  &     ... & 0.93 &  46.30 \\
J101603.14+051302.3$^{*}$  &   ...  &     ... & 5268  &  -13.66 &  ...  &     ... & 0.75 &  46.99 \\
J103351.42+605107.3  &   ...  &     ... & 7607  &  -14.04 &  ...  &     ... & 0.43 &  46.34 \\
J105829.60+013358.8  &   ...  &     ... & 6214  &  -13.48 &  ...  &     ... & 0.61 &  46.38  \\
J115931.84+291443.8  &   ...  &     ... & 5008  &  -13.24 & 4636  &  -13.56 & 0.52 &  46.67 \\
J131028.66+322043.7  &   ...  &     ... & 5278  &  -13.36 & 3633  &  -14.40 & 0.76 &  46.51 \\
J150424.98+102939.1  &  6627  &  -12.90 & 5172  &  -13.35 &  ...  &     ... & 1.69 &  47.44 \\
J155332.69+125651.7$^{*}$  &   ...  &     ... & 3999  &  -13.34 &  ...  &     ... & 1.28 &  47.24  \\
J163515.50+380804.4  &  7452  &  -12.52 & 5144  &  -13.04 &  ...  &     ... & 1.08 &  47.52 \\
J164147.54+393503.3$^{L}$  &   ...  &     ... & 7064  &  -13.44 & 7731  &  -13.72 & 1.84 &  45.11  \\
J222940.09-083254.5$^{*}$  &  7753  &  -12.69 & 3689  &  -13.27 &  ...  &     ... & 1.53 &  46.81 \\
  \noalign{\smallskip}\hline
\end{tabular}
\ec
\tablecomments{0.86\textwidth}{Col. (1): SDSS name. $^{*}$ for
sources with BLR data not available in literatures and newly
measured in this paper. $^{L}$ represents low-confidence association
AGNs. Col. (2): FWHM of broad C IV. Col. (3): flux of broad C IV in
units of $\rm erg~s^{-1}~cm^{-2}$. Col. (4): FWHM of broad Mg II.
Col. (5): flux of broad Mg II in units of $\rm erg~s^{-1}~cm^{-2}$.
Col. (6): FWHM of broad $\rm H\beta$. Col. (7): flux of broad $\rm
H\beta$ in units of $\rm erg~s^{-1}~cm^{-2}$. Col. (8): the
continuum spectral index $\alpha_{\lambda}$ from SDSS spectra. Col.
(9): luminosity at 3000$\rm \AA$.}
\end{table}

\begin{table}\bc
\begin{minipage}[]{100mm}
\caption[]{Black hole mass and BLR
luminosity}\label{tbl5}\end{minipage} \small
 \begin{tabular}{lccccc}
  \hline\noalign{\smallskip}
SDSS Name & log~$M_{\rm BH}$ & log~$L_{\rm BLR}$  & log~$L_{\rm
bol}$ & log~$L_{\rm bol}/L_{\rm Edd}$ & log~$L^{'}_{\rm BLR}$   \\
   & $\rm (M_{\odot})$ & $(\rm erg~s^{-1})$  &  $(\rm erg~s^{-1})$  &  &$(\rm erg~s^{-1})$ \\
  (1) & (2)&(3)&(4)&(5)&(6)\\
  \hline\noalign{\smallskip}
J082447.24+555242.6        &  9.49 & 46.04 & 47.04 &    -0.56 & ...\\  
J092058.46+444154.0        &  9.88 & 47.04 & 48.04 &     0.05 & ...\\  
J094857.33+002225.5        &  8.26 & 44.44 & 45.44 &    -0.93 & 44.39$^{a}$ \\  
J095738.19+552257.7        &  9.20 & 45.11 & 46.11 &    -1.20 & ...\\  
J101603.14+051302.3        &  9.78 & 46.00 & 47.00 &    -0.89 & ...\\  
J103351.42+605107.3        &  9.39 & 45.25 & 46.25 &    -1.25 & ...\\  
J105829.60+013358.8        &  9.25 & 45.32 & 46.32 &    -1.04 & ...\\  
J115931.84+291443.8        &  9.11 & 45.30 & 46.30 &    -0.92 & 45.22$^{a}$ \\  
J131028.66+322043.7        &  8.94 & 45.39 & 46.39 &    -0.66 & 44.72$^{a}$ \\  
J150424.98+102939.1        &  9.50 & 46.36 & 47.36 &    -0.25 & 46.23$^{b}$ \\  
J155332.69+125651.7        &  9.19 & 45.88 & 46.88 &    -0.42 & ...\\  
J163515.50+380804.4        &  9.74 & 46.71 & 47.71 &    -0.14 & 46.53$^{b}$ \\  
J164147.54+393503.3$^{L}$  &  9.19 & 44.80 & 45.80 &    -1.50 & 44.74$^{a}$ \\  
J222940.09-083254.5        &  9.40 & 46.36 & 47.36 &    -0.15 & 46.13$^{b}$ \\  
  \noalign{\smallskip}\hline
\end{tabular}
\ec
\tablecomments{0.86\textwidth}{Col. (1): SDSS name. $^{L}$
represents low-confidence association AGNs. Col. (2): the black hole
mass. Col. (3): BLR luminosity. Col. (4): bolometric luminosity.
Col. (5): the Eddington ratio. Col. (6): BLR luminosity estimated
from single line ($^{a}$ for $\rm H\beta$, $^{b}$ for Mg II).}
\end{table}





\newpage

\begin{figure}

\includegraphics[width=75mm]{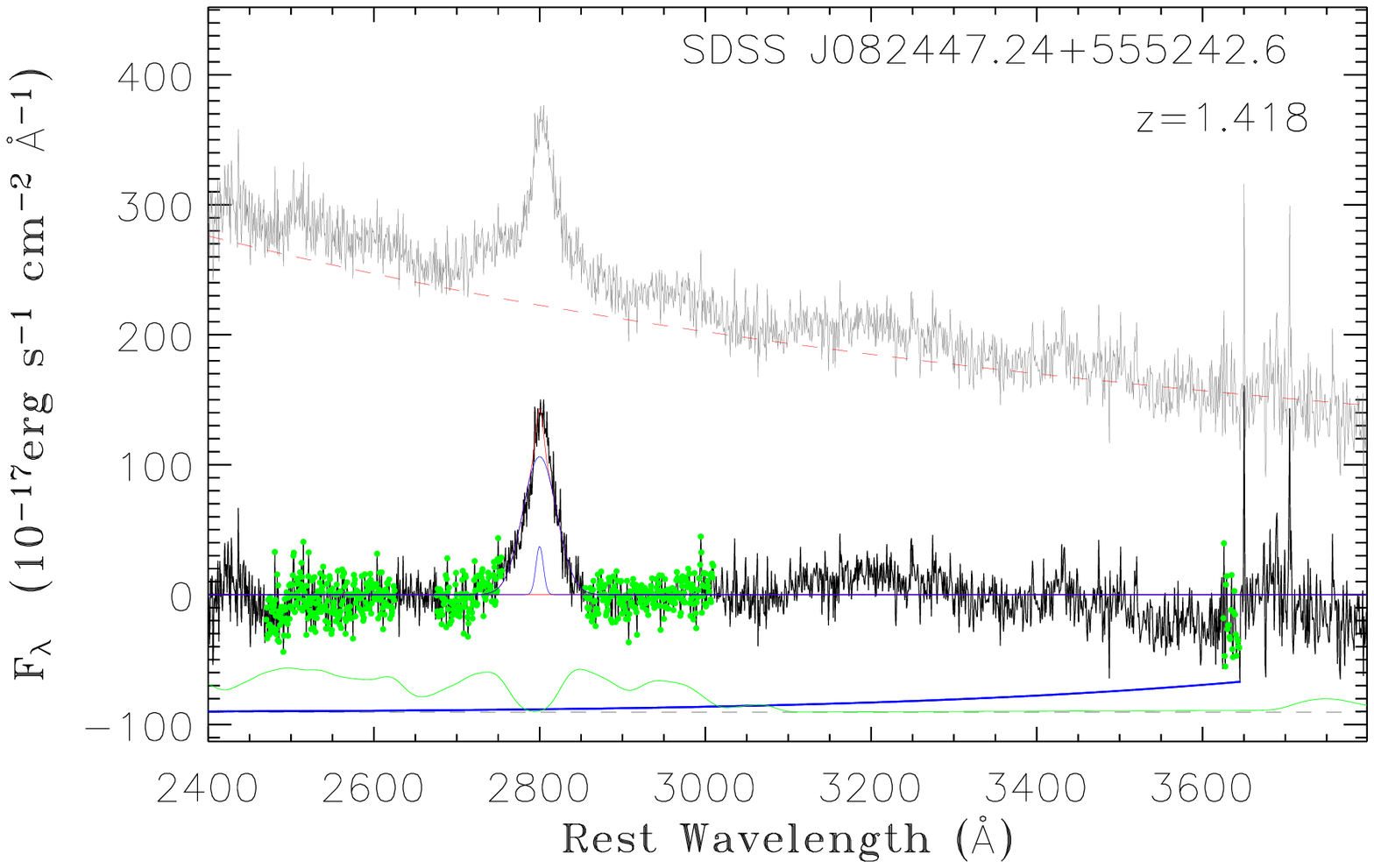}
\includegraphics[width=75mm]{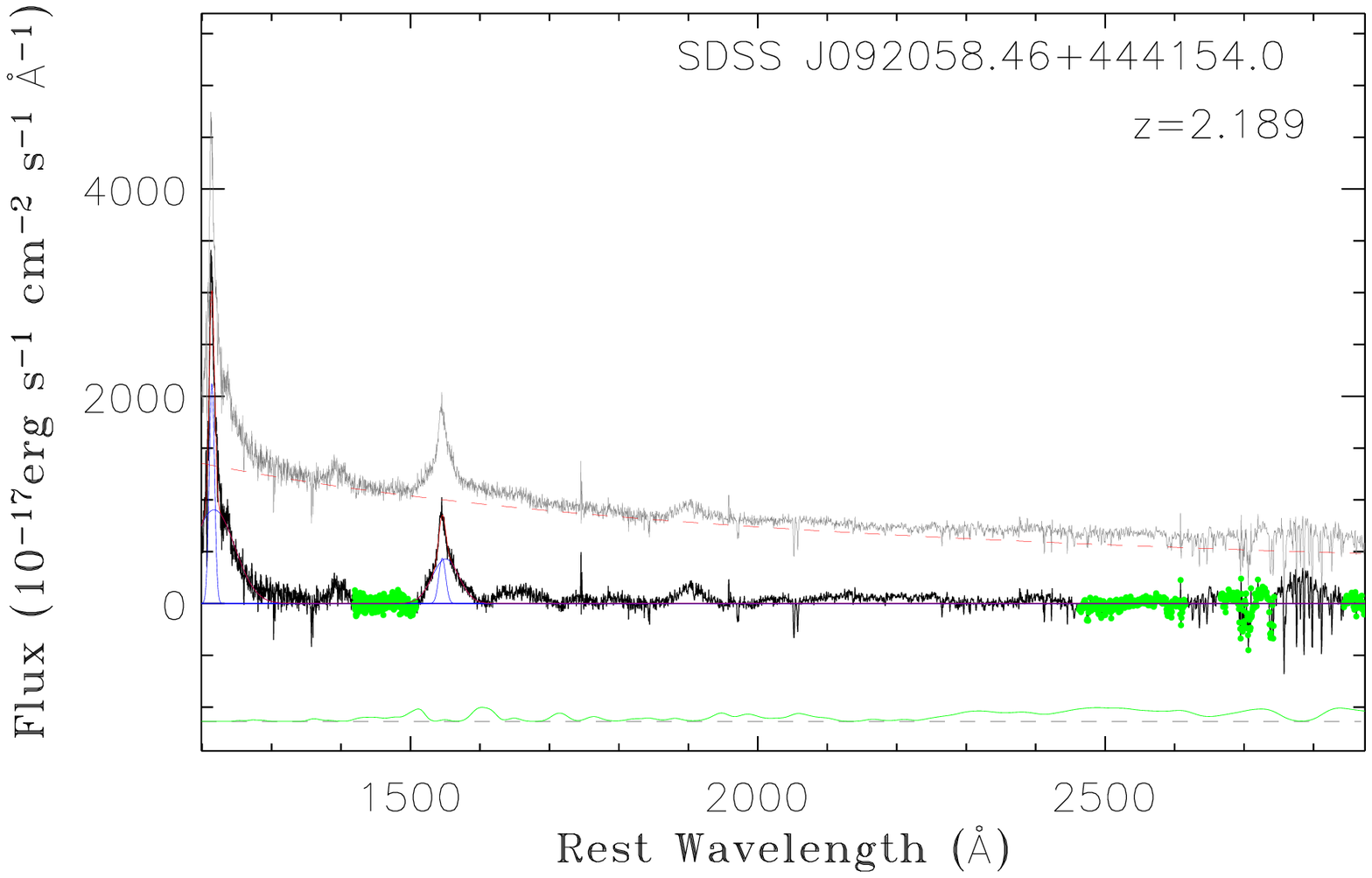}
\includegraphics[width=75mm]{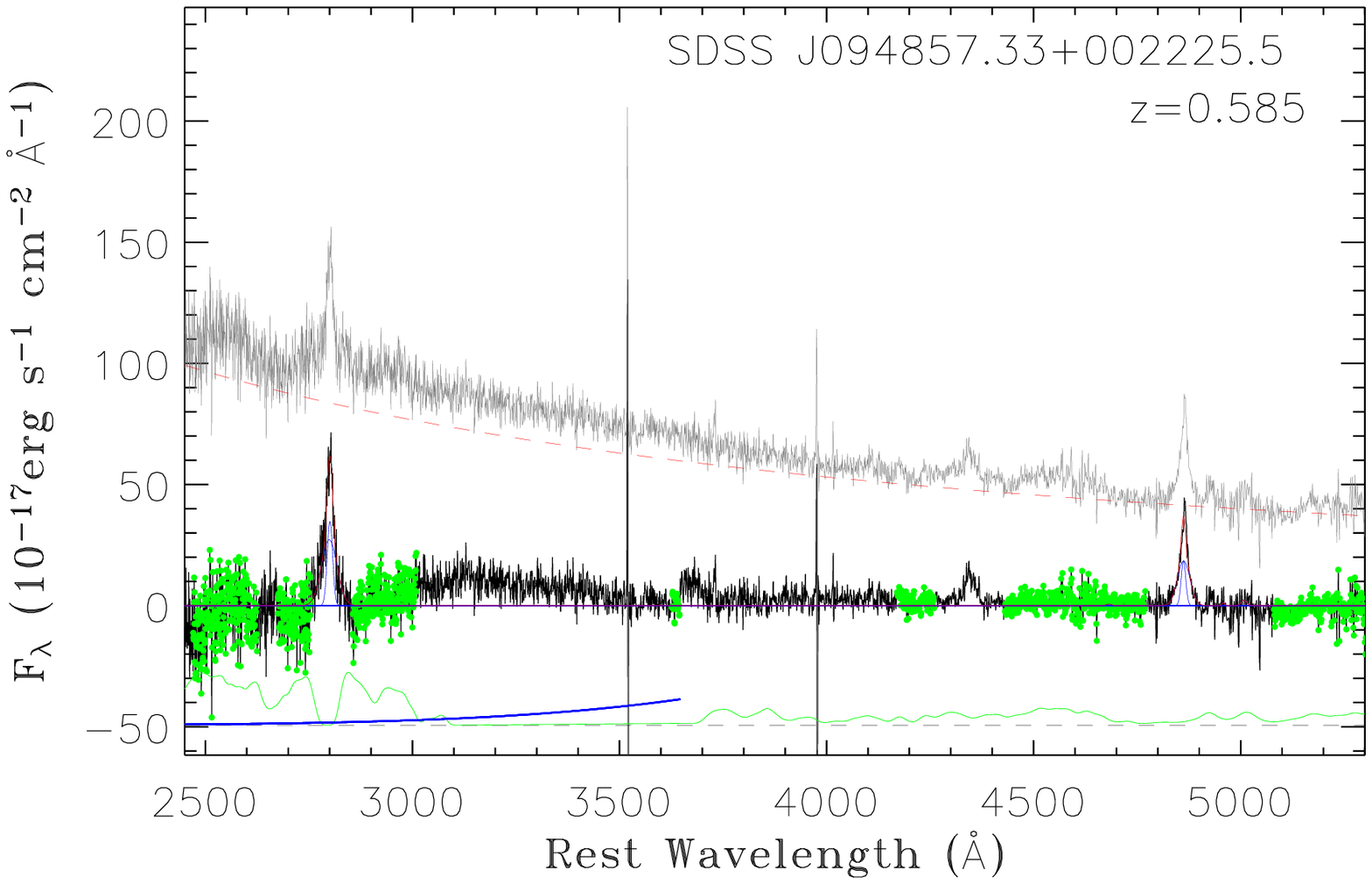}
\includegraphics[width=75mm]{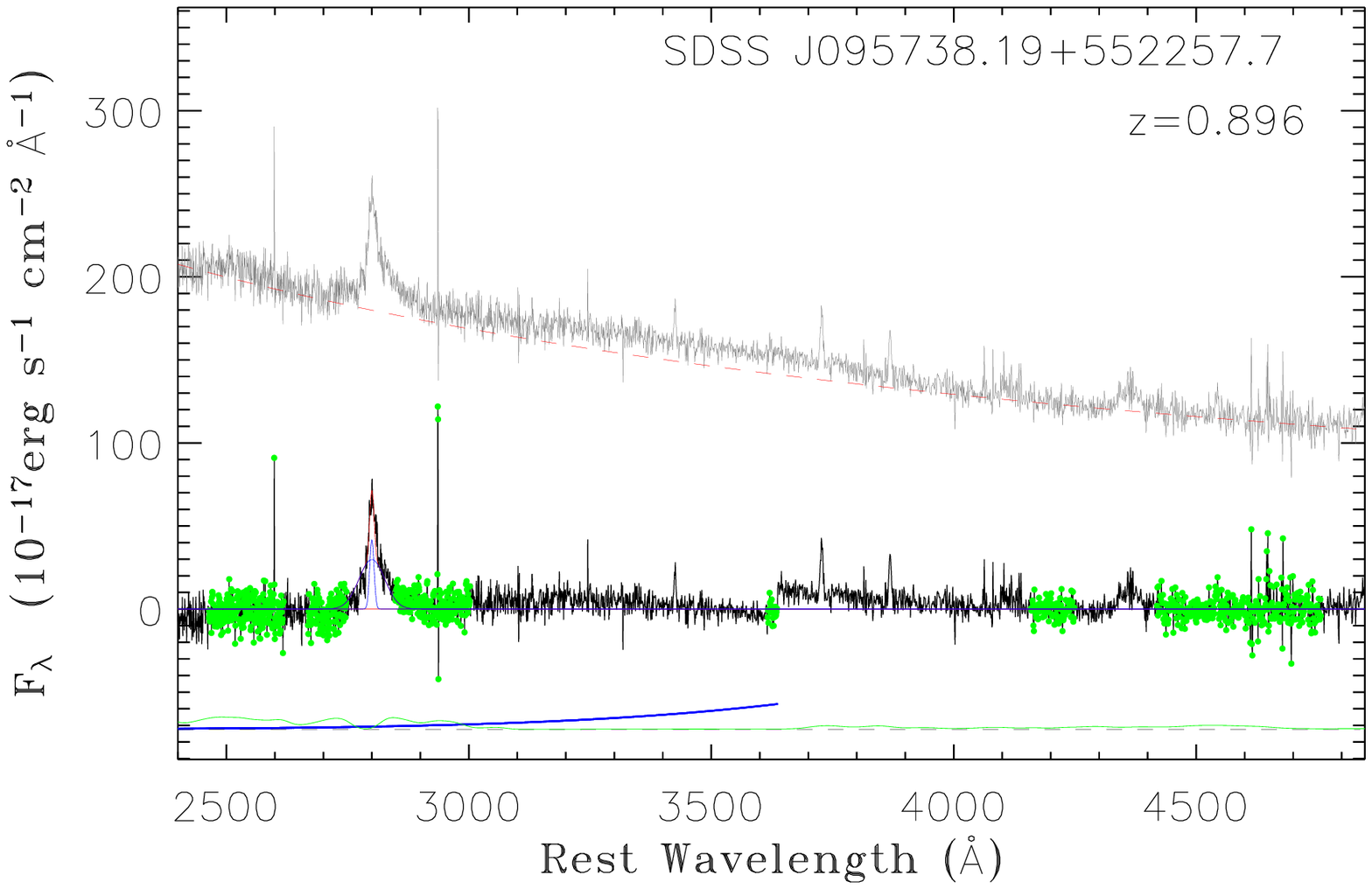}
\includegraphics[width=75mm]{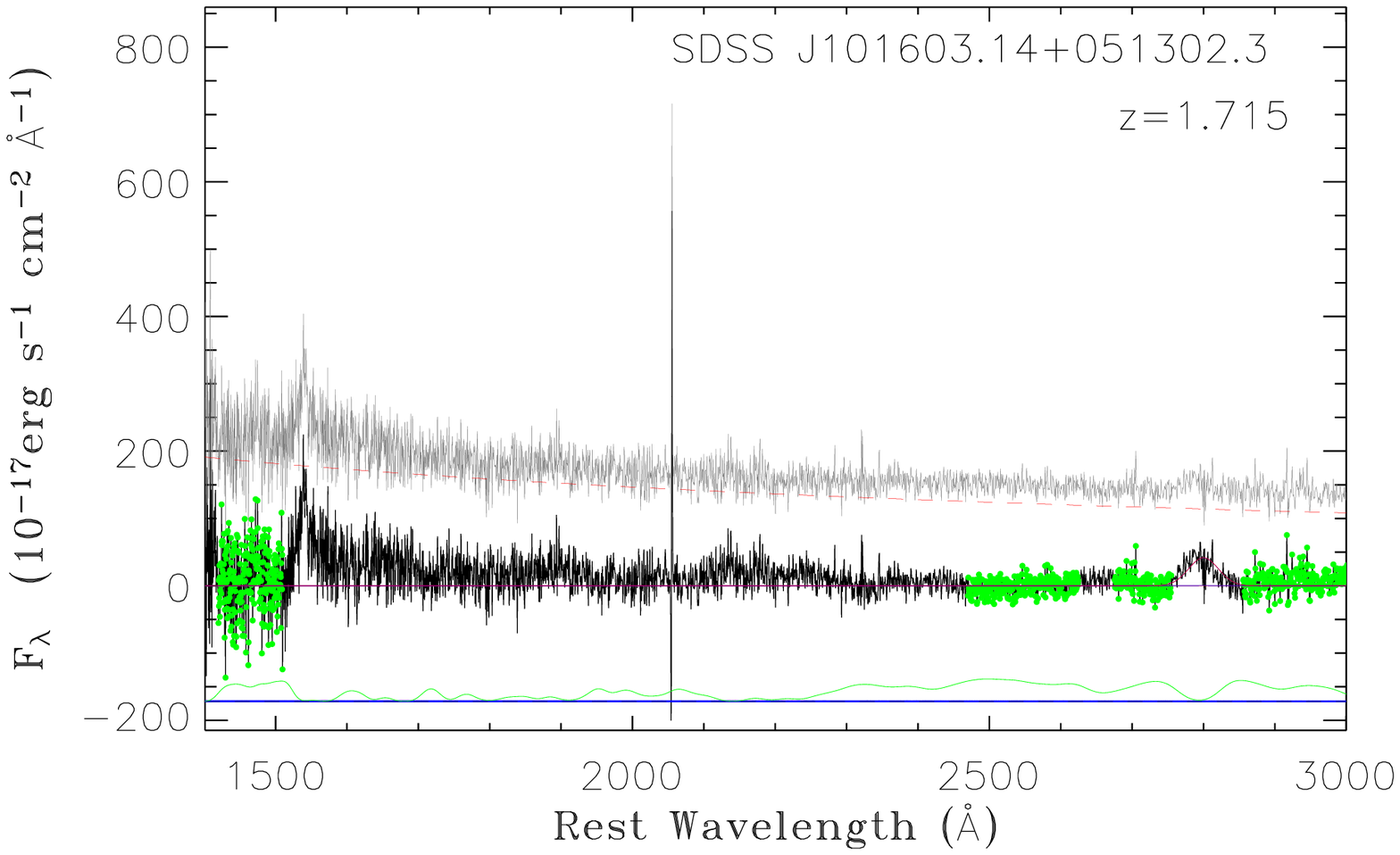}
\includegraphics[width=75mm]{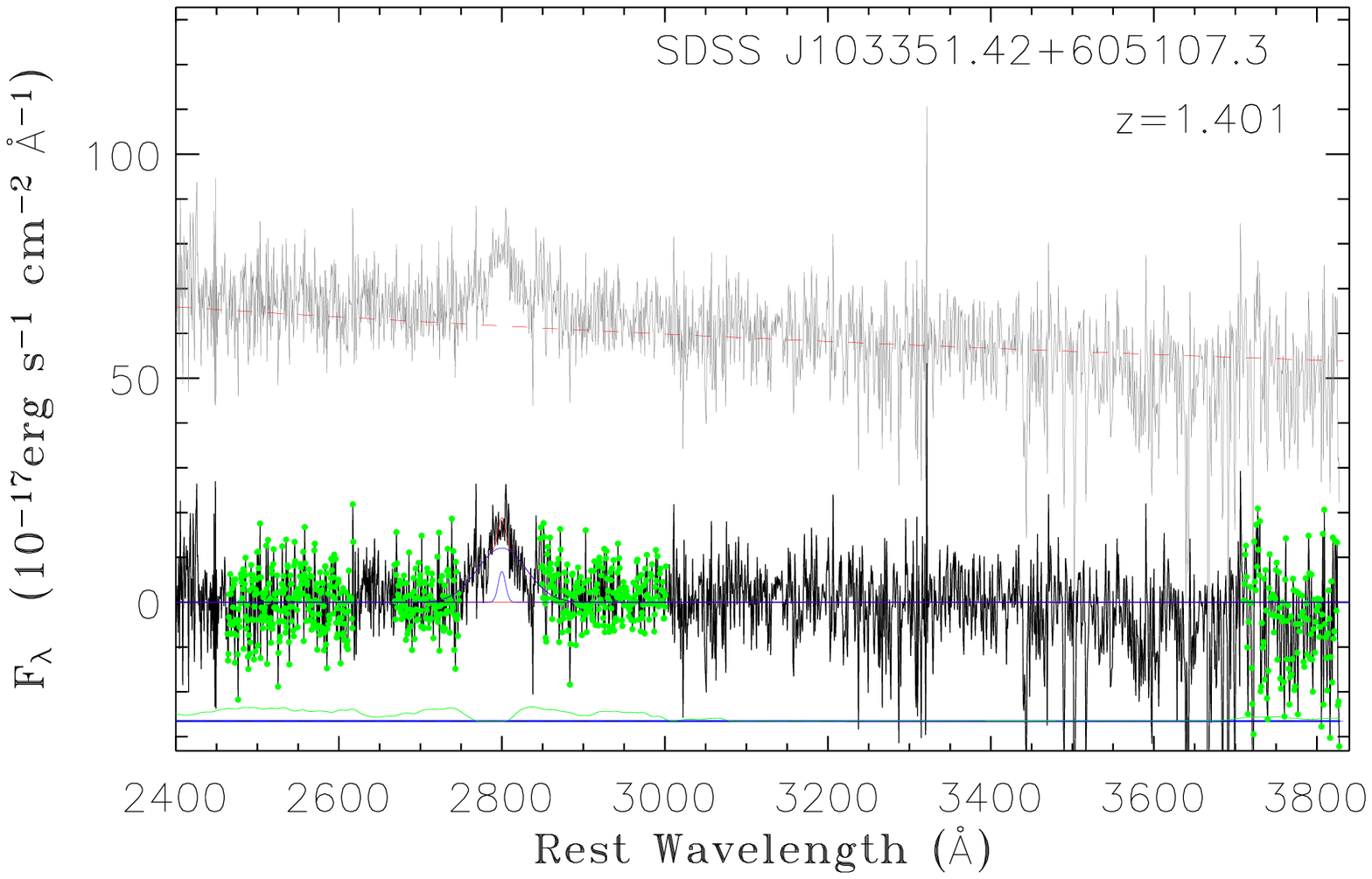}
\includegraphics[width=75mm]{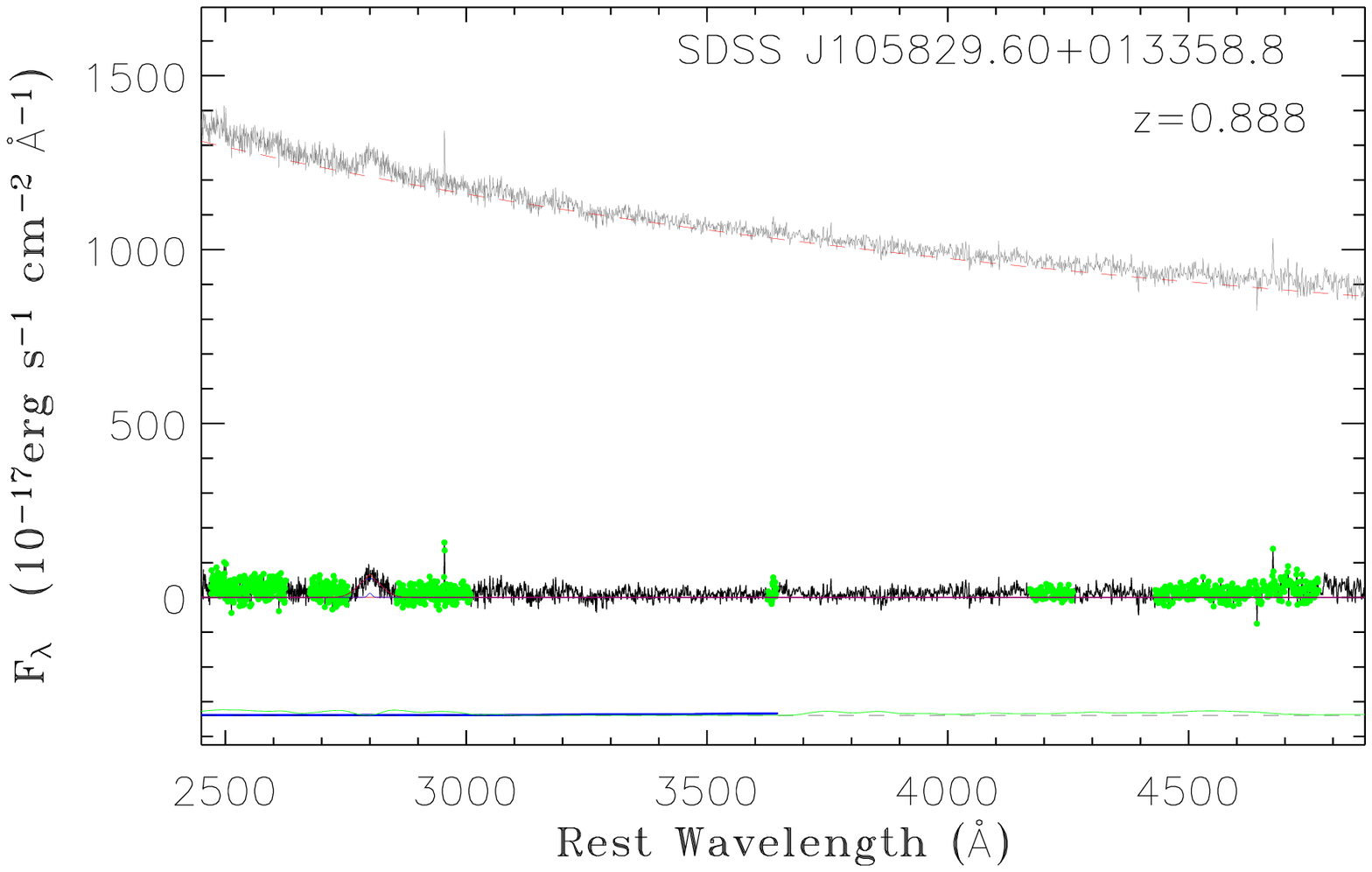}
\includegraphics[width=75mm]{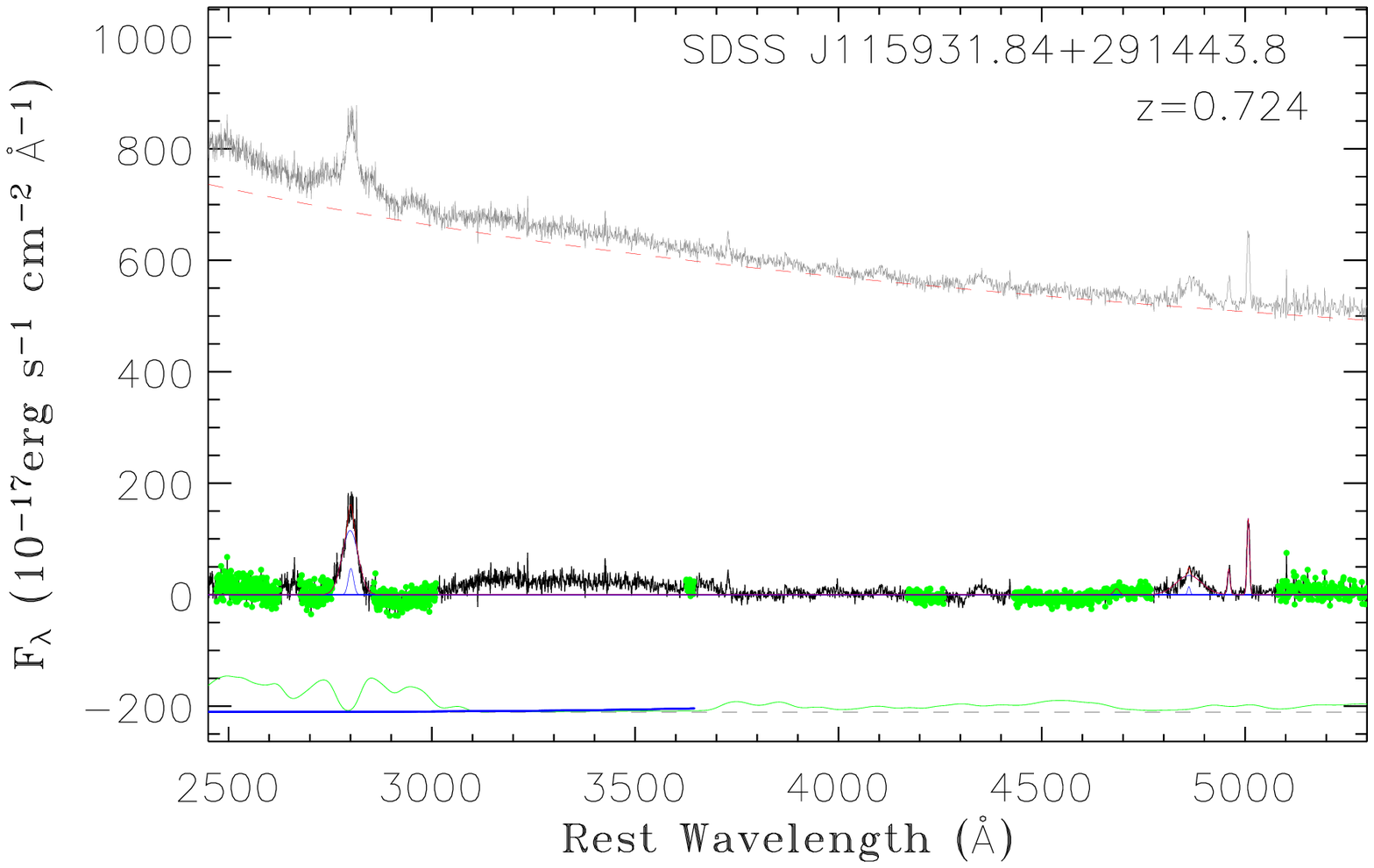}

\caption{The spectral analysis. The source SDSS name and redshift
are presented in each panel in sequences of increasing source R.A..
For each source panel, the top and middle black lines are the
original and the continuum-subtracted spectrum, respectively. The
top dashed red line represents the power-law continuum. The middle
blue lines are the individual line components in multi-line spectral
fitting, and the middle solid red line is integrated lines fitting.
The green spectral region in the continuum-subtracted spectrum is
used to fit Fe II emission and power-law continuum, and the bottom
green line is Fe II emission, which is shifted downwards with
arbitrary unit for the sake of presentation. When necessary, the
Balmer continuum is indicated as solid blue lines in the bottom of
each panel, which begins at the Balmer edge 3646$\rm \AA$.}
   \label{fig1}
\end{figure}

\addtocounter{figure}{-1}

\begin{figure}

\includegraphics[width=75mm]{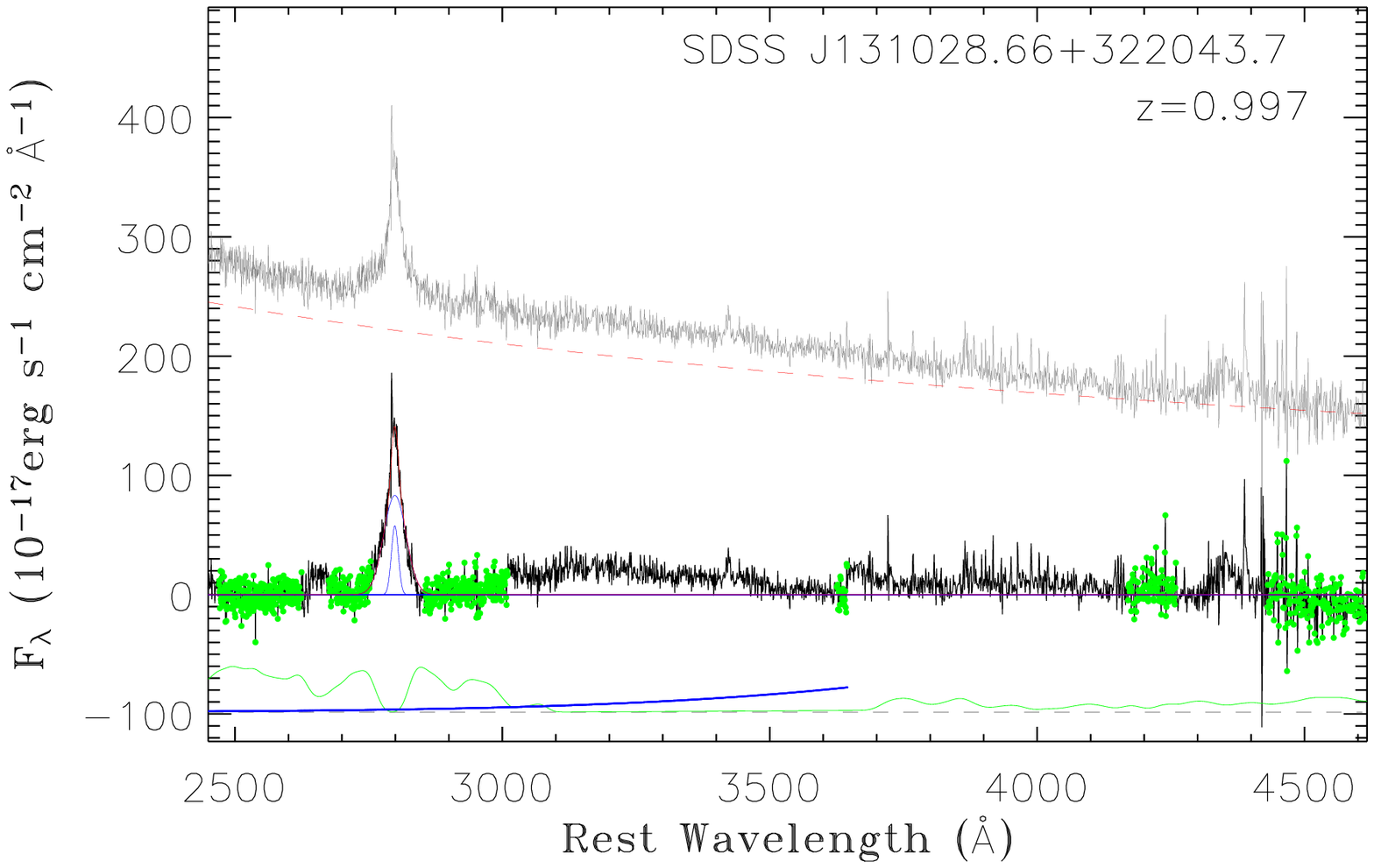}
\includegraphics[width=75mm]{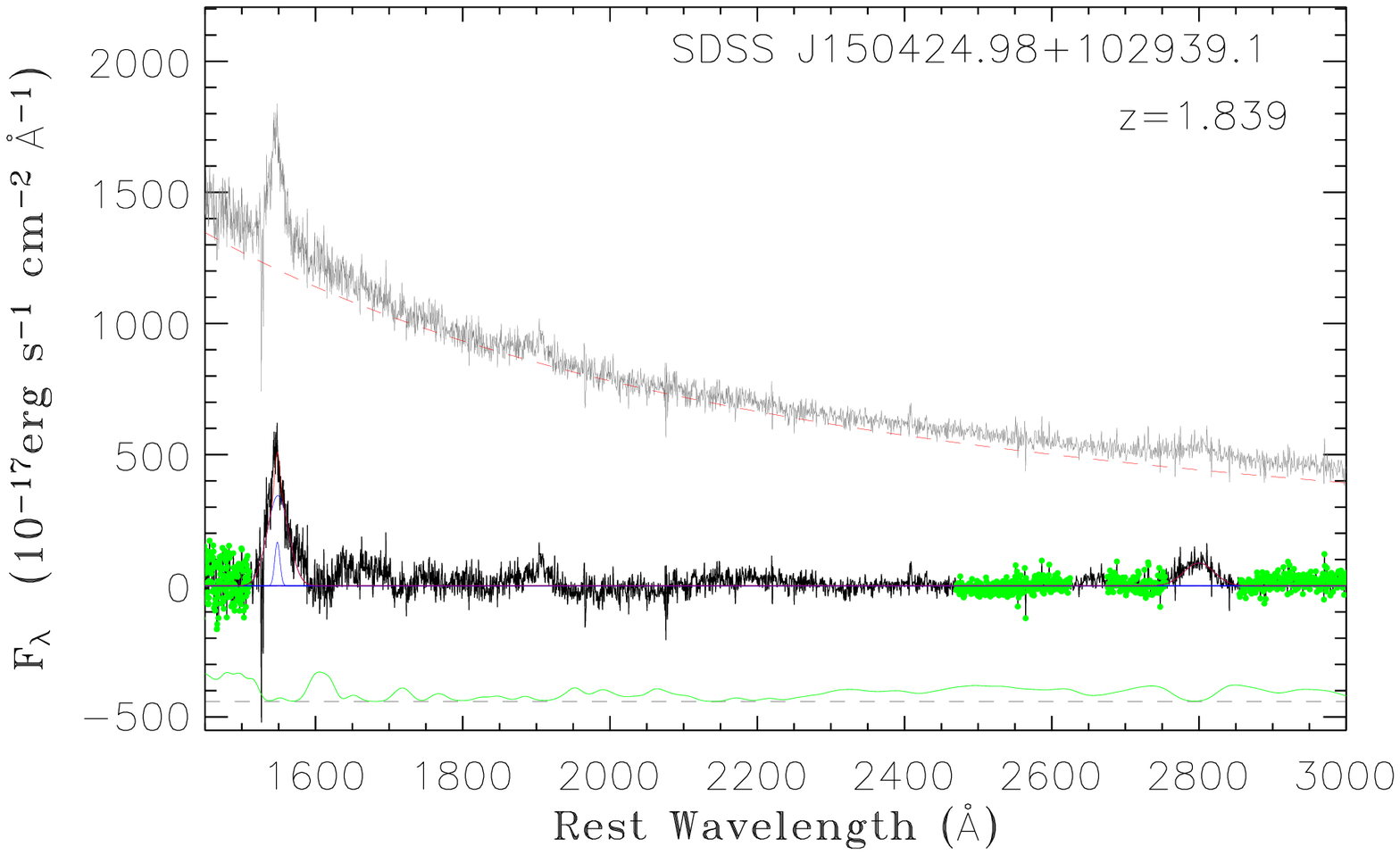}
\includegraphics[width=75mm]{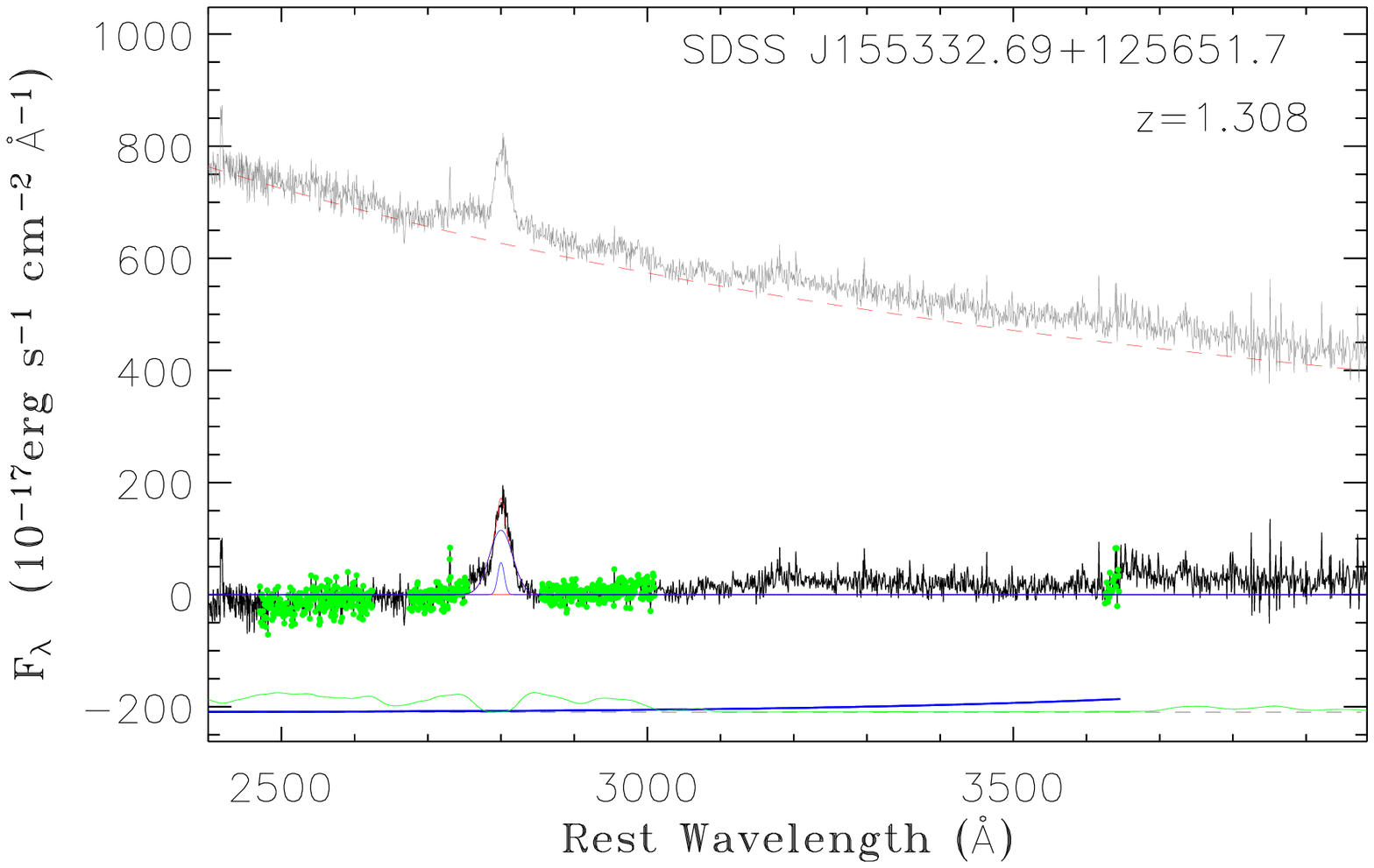}
\includegraphics[width=75mm]{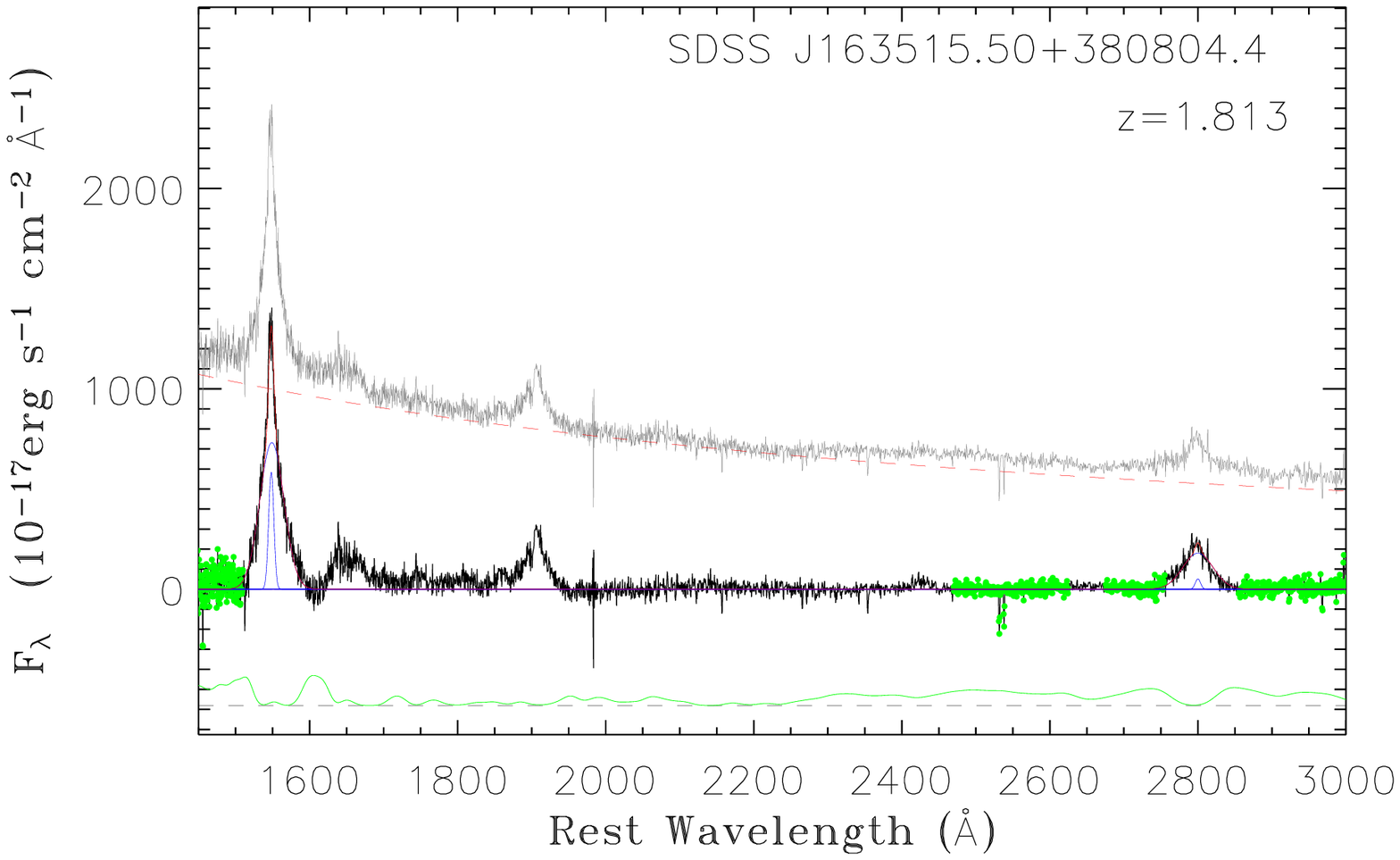}
\includegraphics[width=75mm]{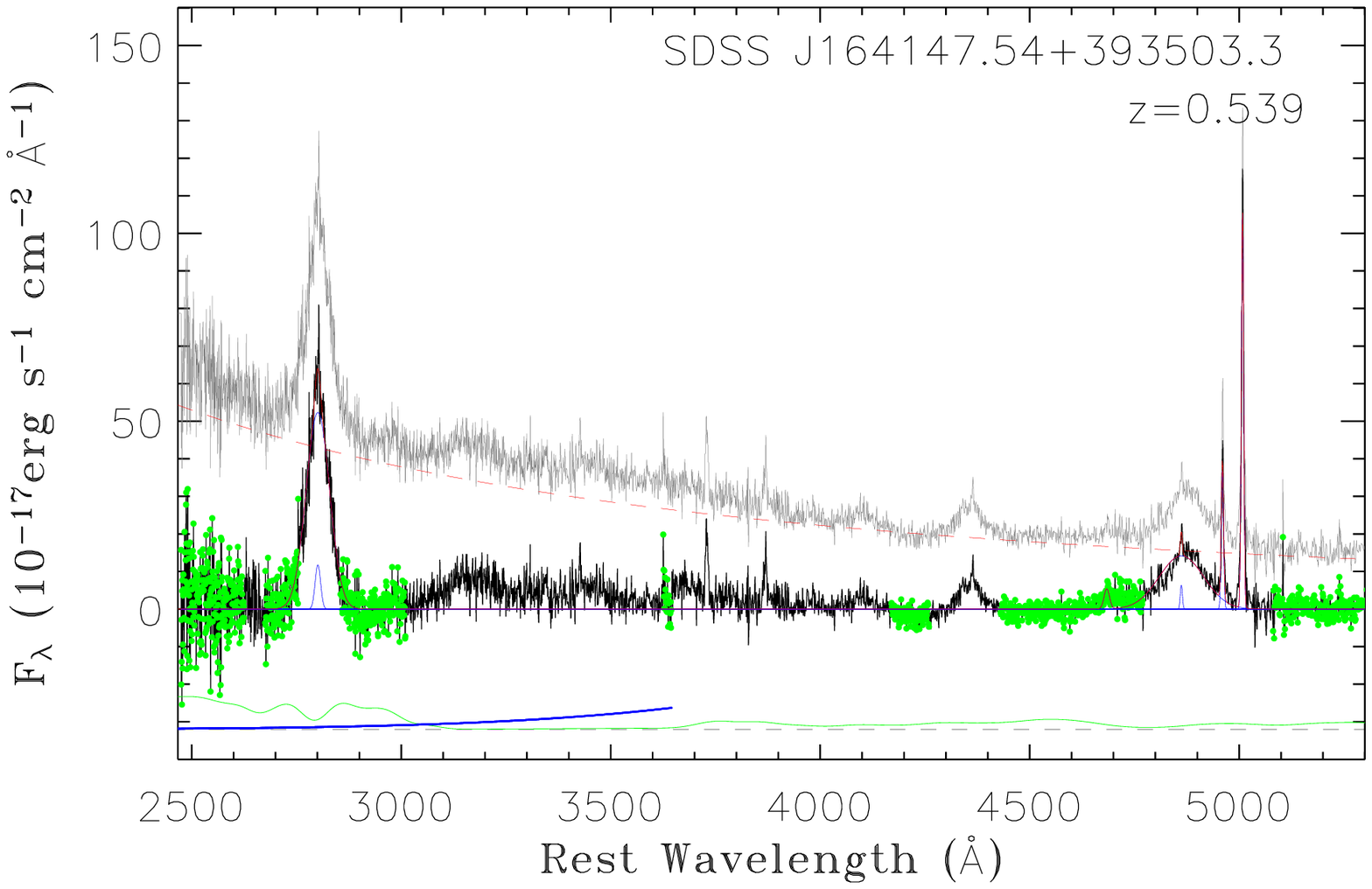}
\includegraphics[width=75mm]{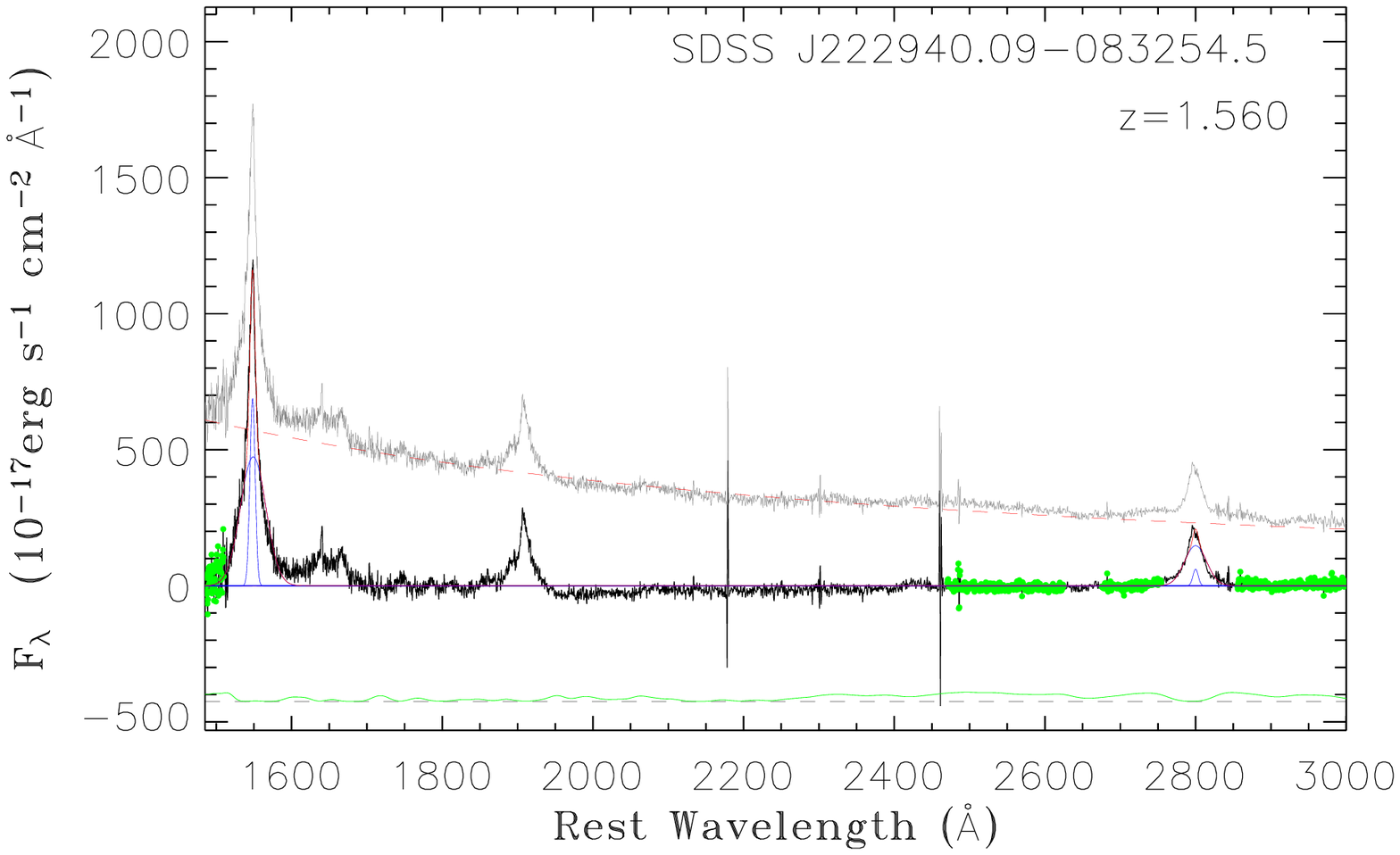}

\caption{continue...}
   \label{fig1}
\end{figure}

   \begin{figure}[h!!!]
   \centering
   \includegraphics[width=12.0cm]{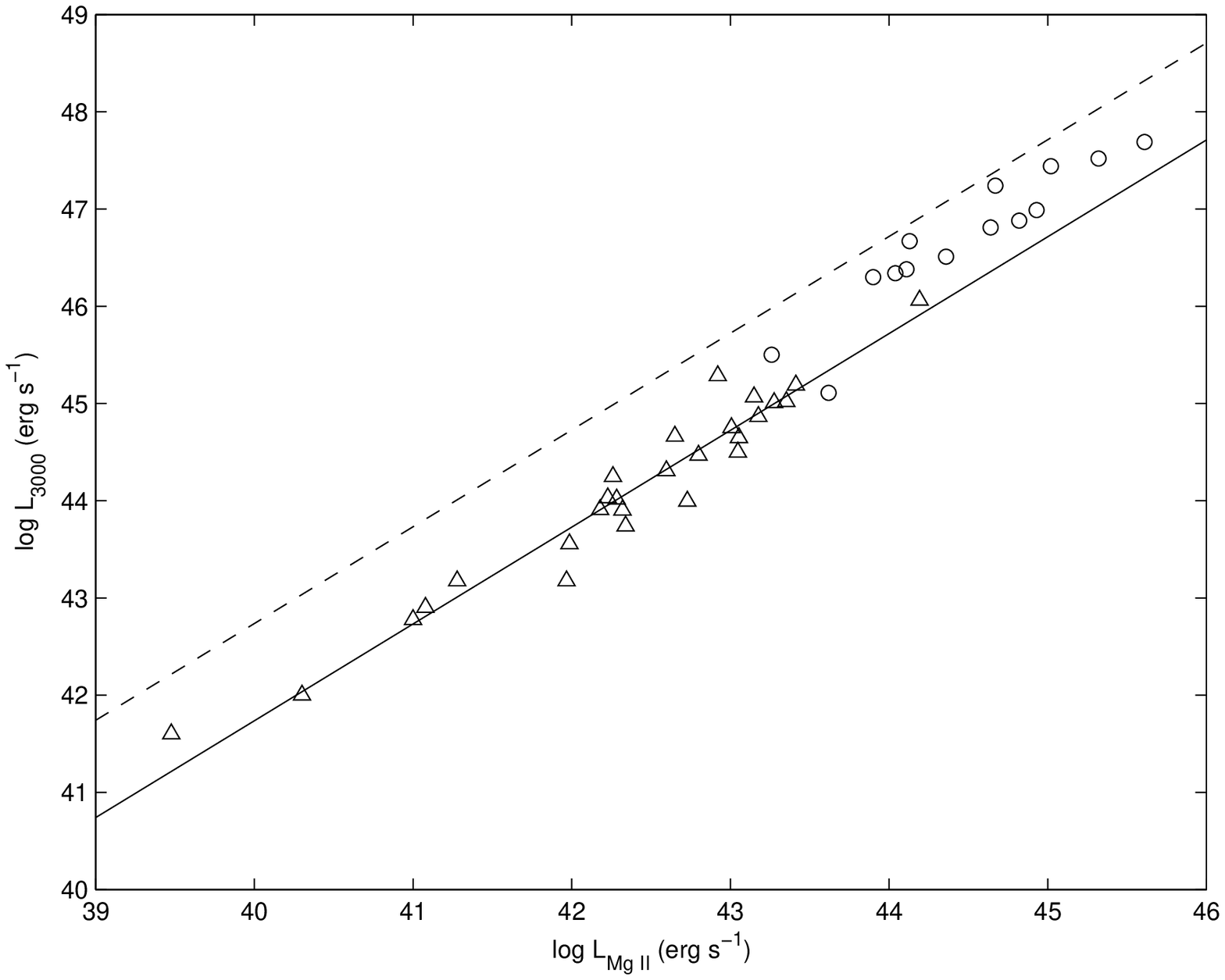}

   \begin{minipage}[]{85mm}

   \caption{Luminosity at $\rm 3000\AA$ versus broad Mg II luminosity.
   The open circles are our FSRQs, while open triangles represent the radio
   quiet AGNs in Kong et al. (2006). The solid
   line is the OLS bisector linear fit to radio quiet
   AGNs in Kong et al. (2006), $\lambda L_{\lambda\rm 3000\AA}=78.5~L_{\rm Mg~ II}^{0.996}$
    (see equation \ref{ml30}). The dashed line represents the deviation from the solid line
    by one order of magnitude in $\lambda L_{\lambda \rm 3000\AA}$, which is used to indicate
    the deviation of our FSRQs from the relation for radio quiet AGNs.
   } \end{minipage}
   \label{fig2}
   \end{figure}

\label{lastpage}

\end{document}